\documentclass[10pt, sigconf]{acmart}

\AtBeginDocument{%
  \providecommand\BibTeX{{%
    \normalfont B\kern-0.5em{\scshape i\kern-0.25em b}\kern-0.8em\TeX}}}

\copyrightyear{2022} 
\acmYear{2022} 
\setcopyright{acmcopyright}\acmConference[ISCA '22]{The 49th Annual International Symposium on Computer Architecture}{June 18--22, 2022}{New York, NY, USA}
\acmBooktitle{The 49th Annual International Symposium on Computer Architecture (ISCA '22), June 18--22, 2022, New York, NY, USA}
\acmPrice{15.00}
\acmDOI{10.1145/3470496.3527395}
\acmISBN{978-1-4503-8610-4/22/06}

\usepackage{algorithm}

\usepackage[shortlabels]{enumitem}
\usepackage{balance}
\usepackage{booktabs}
\usepackage{todonotes}
\usepackage{epsfig}
\usepackage{graphicx}
\usepackage{amsmath}
\usepackage{bm}
\usepackage{lipsum}
\usepackage{subfig}
\usepackage{color}

\usepackage[normalem]{ulem}
\usepackage{bigstrut}
\usepackage{listings}
\usepackage{eso-pic}
\usepackage{tikz}
\usepackage{xspace}
\usepackage{fancyhdr}
\usepackage{mathtools}
\usepackage{soul}
\usepackage{tipa}
\usepackage{textgreek}
\usepackage{siunitx}

\title{Crescent: Taming Memory Irregularities for Accelerating Deep Point Cloud Analytics}

\author{Yu Feng}
\affiliation{
  \institution{University of Rochester}
  \city{Rochester}
  \state{NY}
  \country{USA}
}
\email{yfeng28@ur.rochester.edu}

\author{Gunnar Hammonds}
\affiliation{
  \institution{University of Rochester}
  \city{Rochester}
  \state{NY}
  \country{USA}
}
\email{ghammon5@u.rochester.edu}

\author{Yiming Gan}
\affiliation{
  \institution{University of Rochester}
  \city{Rochester}
  \state{NY}
  \country{USA}
}
\email{ygan10@ur.rochester.edu}

\author{Yuhao Zhu}
\affiliation{
  \institution{University of Rochester}
  \city{Rochester}
  \state{NY}
  \country{USA}
}
\email{yzhu@rochester.edu}

\begin{abstract}

3D perception in point clouds is transforming the perception ability of future intelligent machines. \no{As deep learning-based point cloud algorithms become ever more complicated, an efficient computing system optimized for deep point cloud analytics is increasingly critical. Most of today's hardware architectures are designed and optimized for regular 2D perception kernels such as image/video processing.} Point cloud algorithms, however, are plagued by irregular memory accesses, leading to massive inefficiencies in the memory sub-system, which bottlenecks the overall efficiency.

This paper proposes \proj, an algorithm-hardware co-design system that tames the irregularities in deep point cloud analytics while achieving high accuracy. To that end, we introduce two approximation techniques, approximate neighbor search and selectively bank conflict elision, that ``regularize'' the DRAM and SRAM memory accesses. Doing so, however, necessarily introduces accuracy loss, which we mitigate by a new network training procedure that integrates approximation into the network training process. In essence, our training procedure trains models that are conditioned upon a specific approximate setting and, thus, retain a high accuracy. Experiments show that \proj doubles the performance and halves the energy consumption compared to an optimized baseline accelerator with $< 1\%$ accuracy loss. The code of our paper is available at: \color{blue}{\url{https://github.com/horizon-research/crescent}}.

\end{abstract}



\ifx\figurename\undefined \def\figurename{Figure}\fi
\renewcommand{\figurename}{Fig.}
\renewcommand{\paragraph}[1]{\textbf{#1} }

\newcommand{\Sect}[1]{Sec.~\ref{#1}}
\newcommand{\Fig}[1]{Fig.~\ref{#1}}
\newcommand{\Tbl}[1]{Tbl.~\ref{#1}}

\newcommand{\specialcell}[2][c]{\begin{tabular}[#1]{@{}c@{}}#2\end{tabular}}

\newcommand{\proj}{\textsc{Crescent}\xspace}
\newcommand{\mode}[1]{\underline{\textsc{#1}}\xspace}

\newcommand{\no}[1]{#1}
\renewcommand{\no}[1]{}
\newcommand{\RNum}[1]{\uppercase\expandafter{\romannumeral #1\relax}}

\def\bh{{\mathbf{h}}}
\def\bp{{\mathbf{p}}}

\graphicspath{{figs/}}

\begin{document}
\maketitle



\section{Introduction}
\label{sec:intro}

Recent years have seen an explosive rise of intelligent machines that can perceive, process, and understand visual data. 3D visual data, a.k.a., point clouds, have become increasingly important. \no{Coupled with the prevalence of convenient acquisition devices such as Light Detection and Ranging (LiDARs)~\cite{schwarz2010lidar} and depth cameras, point cloud neural networks have become central to many emerging applications.} Prime examples include localization and mapping in autonomous vehicles~\cite{whitty2010autonomous, liang2018deep} and robotics~\cite{rusu2008towards}, object detection in Augmented and Virtual reality~\cite{stets2017visualization, zhu2017target}, air pollutants detection~\cite{comba2018unsupervised}, and geo-spatial mapping in cultural heritage preservation~\cite{vosselman20013d, guerrero2018pcpnet, openheritage3d}.

Despite much algorithmic development, point cloud networks are inefficient to execute on today's hardware architectures (e.g., GPUs, deep learning/stencil accelerators), most of which are designed and optimized for regular 2D perception domains such as video and image processing~\cite{hegarty2014darkroom, qadeer2013convolution}. Point cloud algorithms, however, exhibit highly irregular computation and memory behaviors and, thus, are ill-suited for architectures built for regular kernels.

The irregularity stems from the fact that memory accesses, which dominate the overall execution efficiency, are input-dependent. As a result, point cloud algorithms exhibit excessive and random (as opposed to streaming) DRAM accesses as well as frequent SRAM bank conflicts that stall the datapath. Many mature optimizations such as tiling, double-buffering, static data layout that are commonly applied to regular kernels such as conventional Deep Neural Networks (DNNs) are either ineffective or not applicable at all.


This paper proposes \proj, which co-designs the algorithm and hardware to tame the irregularities in point cloud algorithms. We start by understanding the sources of memory inefficiency in point cloud algorithms (\Sect{sec:bck}), which points to two main sources. First, point cloud algorithms spend a significant amount of time (up to 80\%)~\cite{feng2020mesorasi} in explicit neighbor searches, which exhibit statically-unknown memory access patterns. Second, the irregular neighbor search necessitates that any subsequent operations must explicitly aggregate points through irregular gather operations instead of simply indexing the memory as in conventional DNNs.

Our key idea is to \textit{impose structures} on memory accesses. We propose an approximate neighbor search algorithm (\Sect{sec:search}) that turns irregular DRAM accesses into streaming accesses. While there are search algorithms that preserve streaming accesses, they often do so at a cost of increasing the search work and/or redundant DRAM accesses by resorting to exhaustive search~\cite{xu2019tigris, pinkham2020quicknn}. We use a different strategy: we use an irregular tree-based algorithm to reduce the search work and selectively elide on-chip bank conflicts to tame the irregularities in tree traversals (\Sect{sec:bank}). This strategy reduces the search work and DRAM traffic by over 40\%.




Our techniques are inexact by nature. Without care, applying them during inference leads to drastic network accuracy loss. To retain accuracy, we propose approximation-aware training (\Sect{sec:train}). Specifically, we integrate the approximation operations into training by modeling hardware behaviors (e.g., bank conflicts) at training time. We show that training with a generic hardware model is usually sufficient, which allows us to avoid tightly coupling training with a particular hardware configuration.

Our training procedure yields models that provide accuracy-vs-performance trade-offs \textit{at inference time without re-training}. This is achieved without increasing the network size or inference overhead. The key is to train a network by sampling not only the input distribution (as with conventional DNN training) but also the distribution of a set of approximation knobs that dictate the accuracy-vs-performance trade-off. In this way, the model's inference is \textit{conditioned} upon a specific approximate setting $\bh$, naturally presenting a different accuracy-vs-performance trade-off for a given $\bh$.

We implement the \proj hardware in a 16nm process node and evaluate it on a set of popular point cloud models. We show that the optimizations introduced in \proj require virtually zero hardware cost and, meanwhile, provide on average 1.9 $\times$ speedup (up to 3.1 $\times$) and 1.5 $\times$ energy reduction (up to 4.2 $\times$) compared to an optimized baseline point cloud accelerator without our optimizations. Notably, the performance and energy gains are achieved with less than 1.0\% accuracy loss.

In summary, this paper makes the following contributions.

\begin{itemize}
    \item We introduce an approximate neighbor search algorithm and its co-designed hardware, which guarantees completely streaming DRAM accesses while reducing the DRAM traffic in point cloud DNNs.
    \item We introduce selectively bank conflict elision, a lightweight scheme to avoid datapath stalls from bank conflicts and reduce SRAM traffic in point cloud networks.
    \item We propose a network training procedure that integrates the approximate neighbor search and selective bank conflict elision into training to mitigate the accuracy loss while providing a flexible accuracy-vs-performance trade-off at inference time.
    \item We show that our optimizations collectively achieve 1.9 $\times$ speedup and 1.5 $\times$ energy reduction for a set of popular point cloud networks compared to a baseline accelerator while sacrificing less than 1\% accuracy.
\end{itemize}

The rest of this paper is so organized. We first characterize the memory inefficiencies, both in DRAM and on-chip SRAM, of today's point cloud networks (\Sect{sec:bck}). We then introduce two techniques to tame the memory inefficiencies: approximate neighbor search that guarantees fully streaming DRAM accesses (\Sect{sec:search}) and selectively bank conflict elision, which streamlines on-chip memory accesses (\Sect{sec:bank}). We then introduce a neural network training procedure that integrates both approximate techniques into the training process to mitigate the accuracy loss (\Sect{sec:train}). After describing the experimental setup (\Sect{sec:exp}), we demonstrate the efficiency of \proj (\Sect{sec:eval}). We then discuss \proj in the broad literature (\Sect{sec:related}) before concluding the paper (\Sect{sec:conc}).
\section{Motivation}
\label{sec:bck}

We first briefly describe the scope of deep point cloud algorithms that this paper targets, and describe the two main algorithmic stages, neighbor search and feature computation, in these algorithms  (\Sect{sec:bck:pc}). We then quantify the memory inefficiencies in both the neighbor search stage (\Sect{sec:bck:search}) and the feature computation stage (\Sect{sec:bck:feature}).

\subsection{Deep Point Cloud Analytics}
\label{sec:bck:pc}

A point cloud is a collection of points, each of which is represented by the [x, y, z] coordinates in the 3D space. Point cloud data are becoming ever more relevant mainly because of two trends: 1) the prevalence of convenient point cloud acquisition devices, e.g., stereo cameras~\cite{lucas1981iterative} and LiDAR~\cite{schwarz2010lidar}, and 2) the emergence of deep learning algorithms that can effectively extract semantics information from point clouds. Today, deep point cloud models are routinely deployed in real-world systems such as Waymo's self-driving cars~\cite{waymolidar} and Google's Augmented Reality toolkit~\cite{arcore}.

\begin{figure}[t]
\centering
\includegraphics[width=\columnwidth]{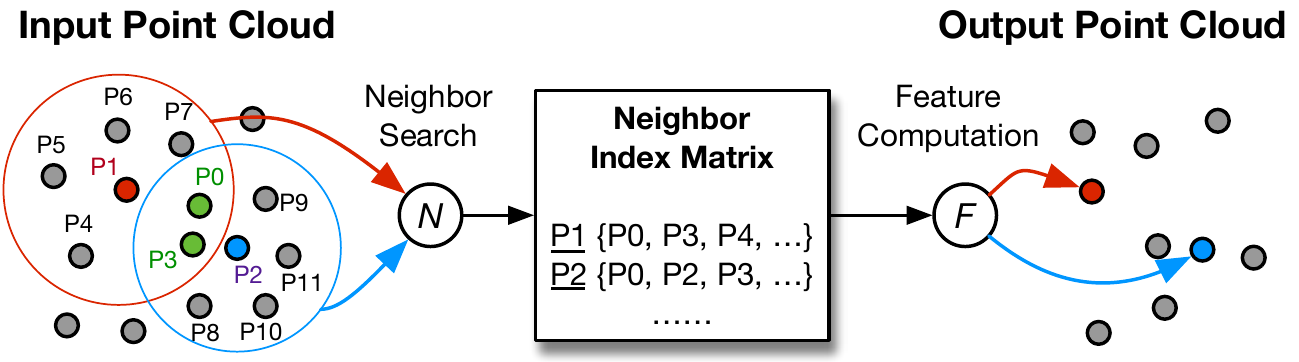}
\caption{A typical layer in a point cloud neural network, which has two main stages: neighbor search and feature computation. Neighbor search in itself is highly irregular as it requires tree traversal and is input-dependent. The feature computation requires irregular memory accesses because the input data are from the neighbor search results.}
\label{fig:pcflow}
\end{figure}

We focus on algorithms that directly operate on raw points, which is by far the most common form of deep point cloud analytics. We refer interested readers to Guo et al.~\cite{guo2020deep} for a comprehensive survey on deep learning for point clouds.

\paragraph{Key Operations} Generally, a point cloud DNN can be abstracted as two stages, as shown in \Fig{fig:pcflow}. Each input point undergoes a neighbor search process. The neighbor search results are stored in a matrix, where each row stores the neighbor indices of a point in the input. The feature computation stage aggregates the neighbors of a point, on which a transformation, usually a Multilayer Perceptron (MLP), is applied, to generate a new output point.

Both stages are important to optimize. A recent study on five popular point cloud networks shows that the execution time ratio of the two stages varies between 1:4 to 4:1~\cite{feng2020mesorasi}, suggesting that neither stage universally dominates.

\subsection{Memory Inefficiencies in Neighbor Search}
\label{sec:bck:search}

Neighbor search in low-dimensional space (e.g., 3D) commonly uses the K-d tree~\cite{bentley1975multidimensional}, which recursively subdivides the search space into two half-spaces using axis-aligned planes. The sub-spaces are organized as a tree, and neighbor search becomes a tree traversal problem. Compared to exhaustive search, the space subdivision strategy is more efficient as it prunes the search space: if the distance of a query $\mathcal{Q}$ and the boundary of a subspace $\mathbb{S}$ is greater than the search radius, all the points in $\mathbb{S}$ can be skipped.

While K-d tree search is inherently parallel (as different search queries are independent), tree traversals are hardware unfriendly. In particular, the memory access patterns are known only at run time, leading to massive inefficiencies in both DRAM and SRAM accesses, which we quantify below.

\begin{figure}[t]
\centering
\begin{minipage}[t]{0.48\columnwidth}
  \centering
  \includegraphics[width=\columnwidth]{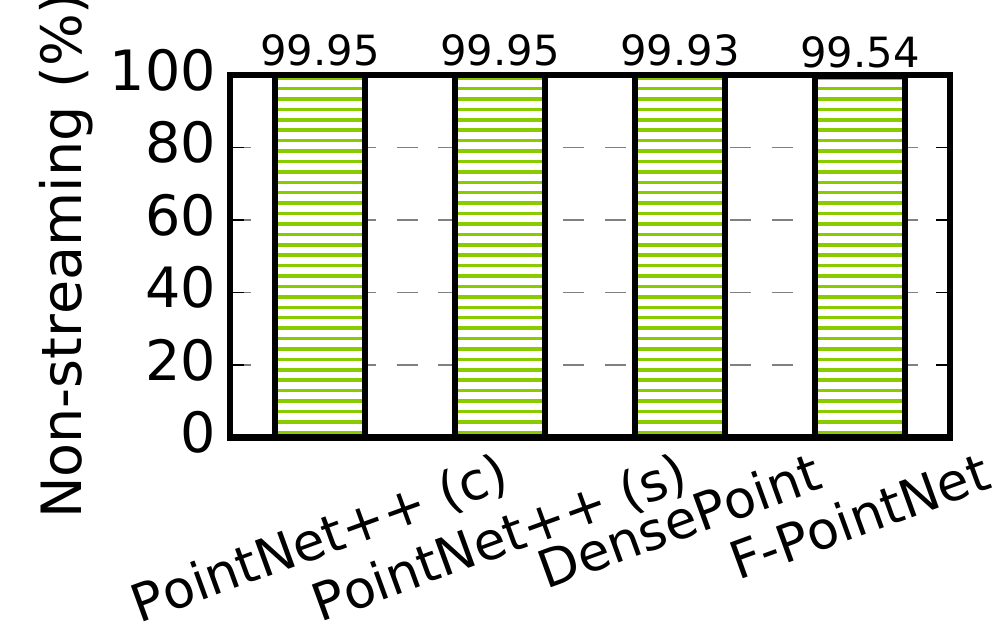}
  \caption{Percentage of non-continuous DRAM accesses in common point cloud networks.}
  \label{fig:dram_non_stream}
\end{minipage}
\hspace{2pt}
\begin{minipage}[t]{0.48\columnwidth}
  \centering
  \includegraphics[width=\columnwidth]{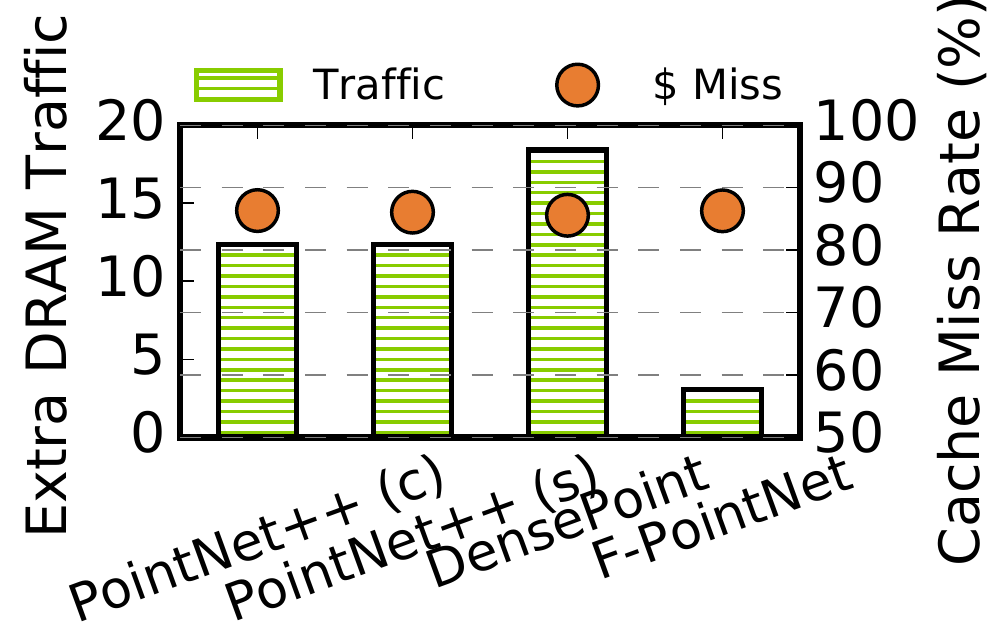}
  \caption{Ratio of actual DRAM traffic vs. the theoretical minimum and cache miss rate in neighbor search.}
  \label{fig:dram_needed_data}
\end{minipage}
\end{figure}

\paragraph{DRAM} DRAM access inefficiency in neighbor search is manifested in two ways: non-streaming accesses and redundant accesses.
The DRAM accesses are non-streaming because the inputs (points) are arbitrarily distributed in the search space. If two queries being processed in parallel are spatially far-apart, they will likely exercise different parts of the K-d tree that are non-contiguous in memory. Even within the same query, tree nodes consecutively accessed during traversals are likely non-continuous in memory due to the control-flow heavy nature of tree traversal. \Fig{fig:dram_non_stream} shows the percentage of non-continuous DRAM accesses across four popular point cloud DNNs (see \Sect{sec:exp} for a comprehensive experimental setup). Almost all DRAM accesses are non-continuous.

The non-streaming nature coupled with large point cloud data size leads to redundant DRAM accesses. For instance, in the popular KITTI dataset~\cite{geiger2012we}, the total points and queries in a typical scene \textit{alone} can be over tens of MBs (not considering the network weights, activations, etc.), larger than what a mobile SoC can accommodate. Thus, points are loaded on-chip in chunks (analogous to tiling in conventional DNNs). Since not all data in each chunk will be used when they are loaded due to the non-streaming access pattern, a great amount of DRAM accesses are wasted.

\Fig{fig:dram_needed_data} quantifies the excessive DRAM accesses and cache miss rate in neighbor search. The left $y$-axis shows the ratio of the amount of DRAM requests (in bytes) to the actual data theoretically needed by the algorithm (i.e., reading each query and search point once). The data are obtained by simulating an unrealistic 10~MB fully-associated cache running a neighbor search on a typical KITTI-constructed scene with about 1.2 million points. Even with this unrealistic SRAM structure, searches in many models have about 10$\times$ more DRAM traffic than what is strictly required. Realistic mobile accelerators would allocate an even smaller buffer for neighbor search to accommodate other data structures such as DNN weights and activations. The right $y$-axis quantifies the corresponding cache miss rates, which are over 85\%.


\paragraph{SRAM} The on-chip memory accesses in K-d tree search are also inefficient because of the frequent bank conflicts. In regular kernels such as stencil pipelines~\cite{whatmough2019fixynn, qadeer2013convolution} where the memory access pattern is statically known, one could carefully interleave data in the SRAM banks to avoid bank conflicts~\cite{zhou2021characterizing, kirk2016programming}. In contrast, on-chip memory accesses in neighbor search are input-dependent and, thus, bank conflicts are inevitable.

\begin{figure}[t]
\centering
\begin{minipage}[t]{0.48\columnwidth}
  \centering
  \includegraphics[width=\columnwidth]{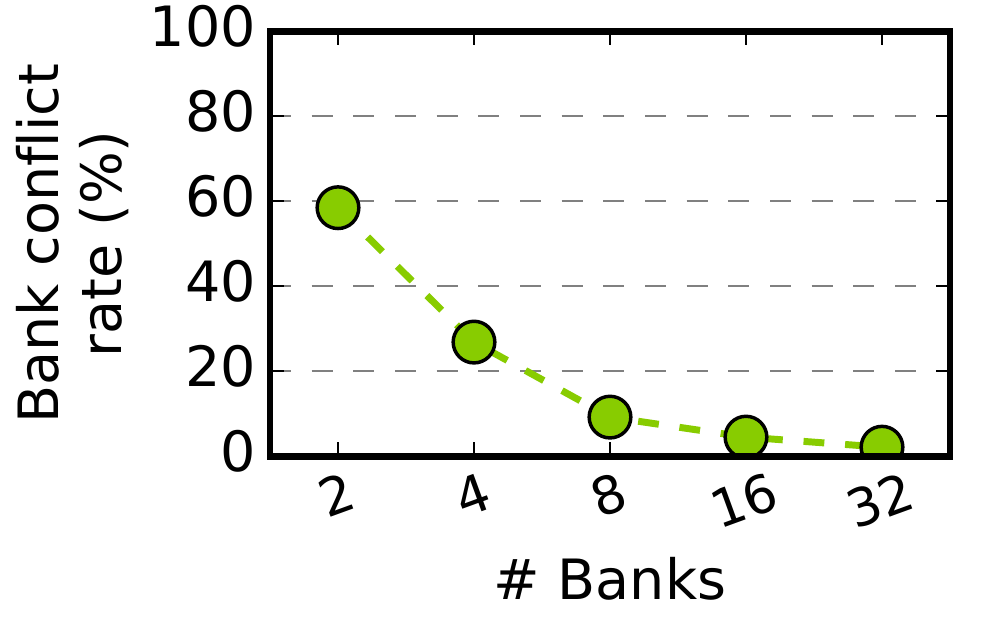}
  \caption{Neighbor search bank conflict rate in Pointnet++(c) vs. the number of banks under 8 concurrent queries.}
  \label{fig:sram_conflict}
\end{minipage}
\hspace{2pt}
\begin{minipage}[t]{0.48\columnwidth}
  \centering
  \includegraphics[width=\columnwidth]{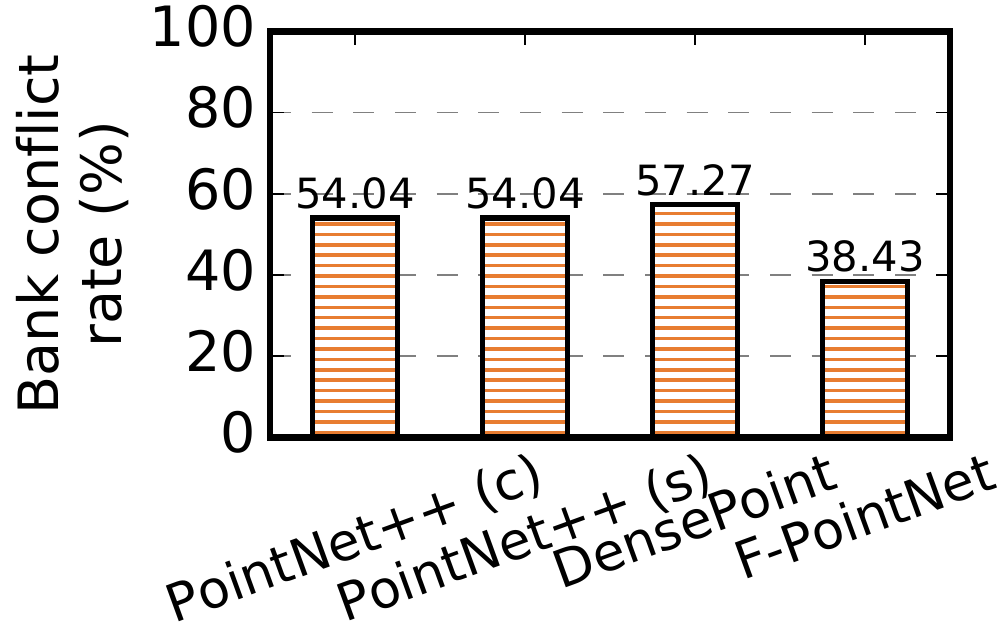}
  \caption{SRAM bank conflict rate in aggregation, assuming 16 banks and 16 concurrent memory requests.}
  \label{fig:group_sram_conflict}
\end{minipage}
\end{figure}

\Fig{fig:sram_conflict} quantifies the bank conflicts by showing the percentage of SRAM accesses that are bank-conflicted and how the percentage varies with the number of banks. We assume an unrealistically large 10~MB buffer and 8 concurrent SRAM requests. With 4 banks the bank conflict rate is 26.9\%. The bank conflict rate is reduced to 2.1\% only when the number of banks quadruples the number of simultaneous requests.

Using a heavily-banked SRAM design is highly undesirable. A large number of banks requires a more costly crossbar design~\cite{agarwal2009garnet, grot2011kilo}, as the crossbar area grows quadratically with the number of banks. Using an Arm memory compiler~\cite{artisanmc}, we find that the crossbar area is twice as much as the memory arrays under a 32-bank configuration. In addition, a higher bank count also reduces the memory array size, which increases the per-bank overhead (peripheral circuits, BIST, redundancy)~\cite{weste2015cmos}.



\subsection{Memory Inefficiencies in Feature Computation}
\label{sec:bck:feature}

Unlike neighbor search, the DRAM accesses in the feature computation stage are completely streaming. The on-chip memory accesses, however, are met with frequent bank conflicts.

Feature computation is broken down into two steps: 1) aggregate the neighbors for each input point $\bp_i$ using the neighbor indices generated in the neighbor search stage, and 2) compute an output point $\bp_o$ from each $\bp_i$ by applying a function, usually a MLP, to the neighbors of $\bp_i$. Step 2 is accelerated on today's DNN accelerators.

Step 1 is analogous to fetching data from the input feature map in a conventional DNN. However, conventional DNNs access consecutive feature map elements with statically-known patterns. Therefore, a compiler lays out data in the SRAM such that a simple single-bank, single-port memory array (using wide words) could serve memory requests from tens or hundreds of PEs in one cycle without stalling the PEs~\cite{jouppi2017datacenter, armmlproc, zhou2021characterizing}.

However, point cloud networks access non-consecutive memory in this step, because the neighbors of a point can be arbitrary. Therefore, the SRAM serving points are usually banked. Worse, the access pattern is statically-unknown, as it depends on the neighbor search results, which, in turn, depend on the inputs. Therefore, bank conflicts are inevitable.

\Fig{fig:group_sram_conflict} quantifies the severity of bank conflicts in point aggregation by showing the percentage of SRAM accesses that are bank-conflicted in aggregating the points. We assume a 16-bank SRAM design with a total size of 64 KB. Across the four models, the bank conflict rate is at least 38.43\% and can be as high as 57.27\%. Increasing the number of banks is undesirable as it requires a more costly crossbar and/or a higher per-bank overhead due to the smaller memory arrays~\cite{weste2015cmos}.

\section{Fully-Streaming Neighbor Search Algorithm}
\label{sec:search}

We introduce our neighbor search algorithm and explain how it fundamentally improves the DRAM access efficiency by allowing completely streaming memory accesses (\Sect{sec:search:algo}). We then describe the co-designed neighbor search hardware (\Sect{sec:search:hw}). Finally, we discuss the key knob in our algorithm that dictates the accuracy-vs-performance trade-off (\Sect{sec:search:approx}).

\begin{figure}[t]
\centering
\includegraphics[width=\columnwidth]{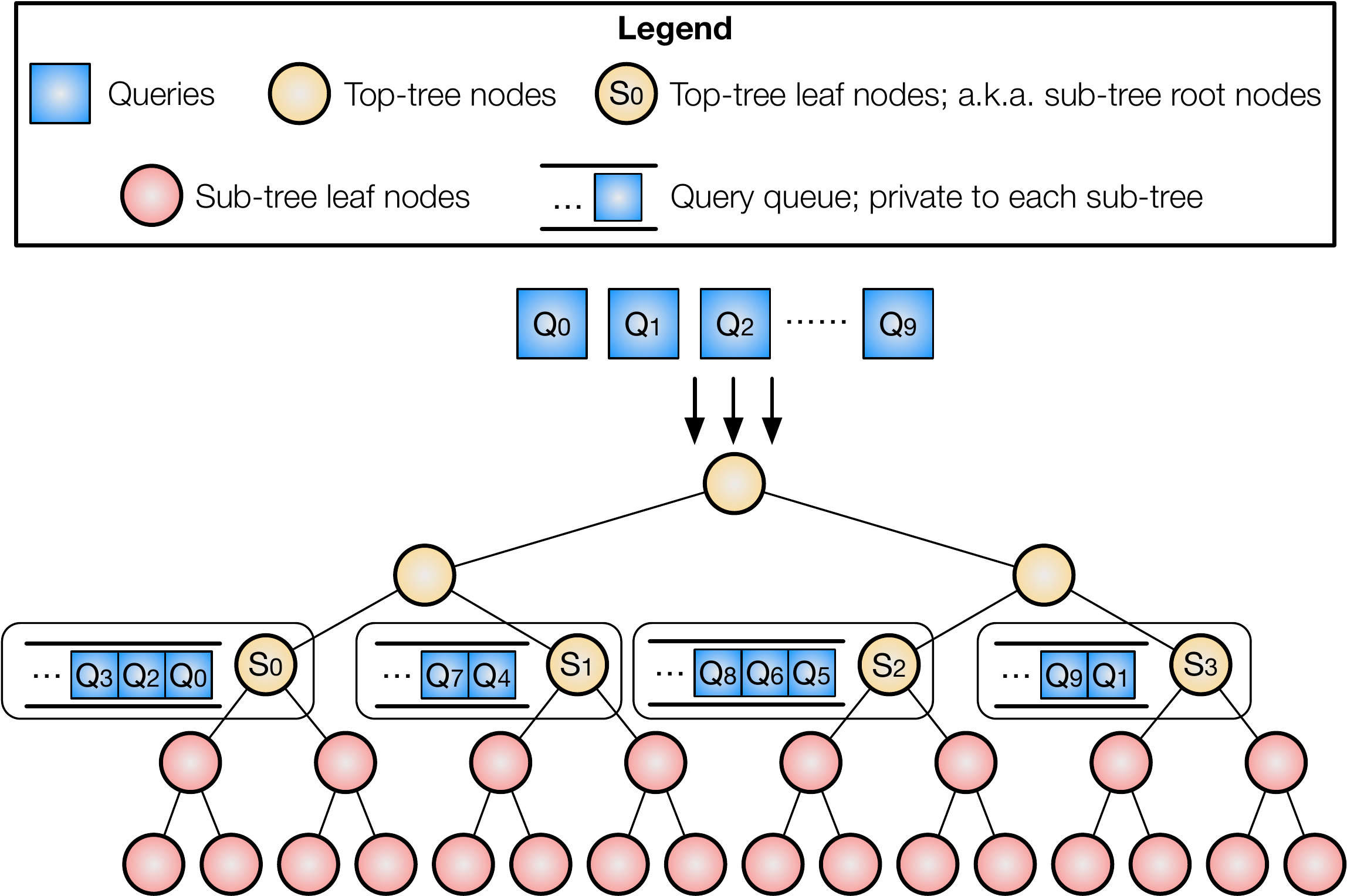}
\caption{The two-level tree data structure of our neighbor search algorithm. In the first stage, queries traverse the top-tree and are assigned to a particular sub-tree in the end. In the second stage, queries search neighbors in their assigned sub-tree, and backtracking is limited to within the sub-tree.}
\label{fig:algo}
\end{figure}

\begin{figure*}[t]
\centering
\includegraphics[width=2\columnwidth]{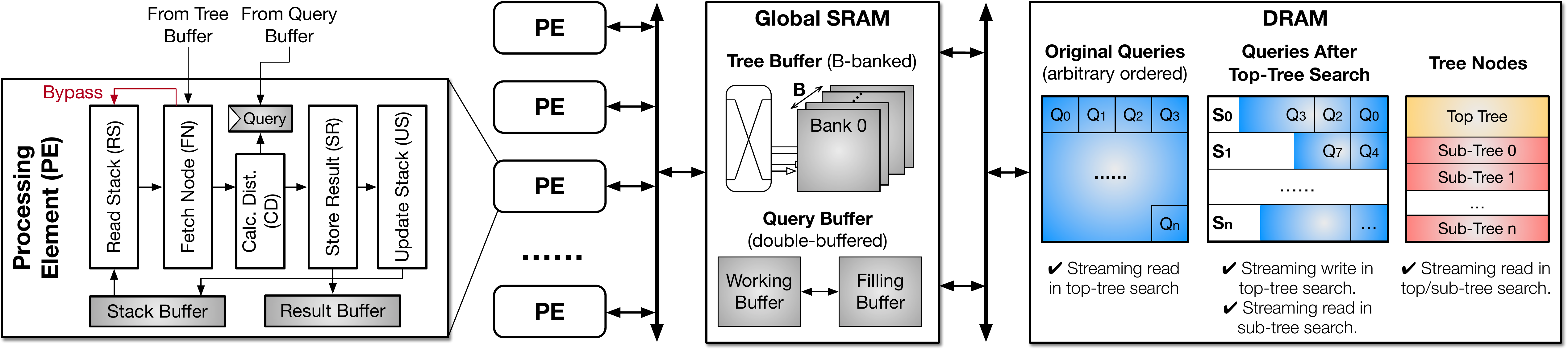}
\caption{Neighbor search hardware engine, which enables a fully-streaming access to DRAM. The same hardware is used for both top-tree search and for the sub-tree searches, simplifying the hardware design.}
\label{fig:nsarch}
\end{figure*}

\subsection{Algorithm}
\label{sec:search:algo}

Our algorithm splits the K-d tree into a top tree and a set of sub-trees. Each top-tree leaf node is also the root node of a sub-tree. The search is then naturally divided into two stages: a top-tree search stage and a sub-tree search stage. The two stages themselves are massively parallel but are serialized with each other. \Fig{fig:algo} illustrates the idea.

In the first stage, all the queries search the top-tree (a binary search tree) until they reach the leaf nodes of the top-tree, at which point the queries are assigned to the corresponding sub-trees. Conceptually, each sub-tree has a queue that stores all the incoming queries. At the end of the first stage, queries in the sub-tree queues are written back to the memory in preparation for the second stage. In actual hardware, a queue has a fixed size. Thus, the store back to the memory is phased, as we will discuss later.

Once all the queries finish the first stage, the algorithm enters the second stage, where queries in each sub-tree are searched against the corresponding tree. For each sub-tree, the search process is exactly the same as that in the top-tree with a critical difference: queries are allowed to backtrack when they reach a leaf node of the sub-tree. This is necessary for a query to find all its neighbors.

However, we limit the backtracking to the sub-tree. The intuition is that nodes in other sub-trees are naturally far away from the query and thus are less likely to be neighbors. Architecturally, this ensures that each sub-tree and each query is loaded to SRAM once --- at a cost of accuracy loss. We will discuss the accuracy implication of this design decision in \Sect{sec:search:approx} and how to mitigate the accuracy loss through approximation-aware network training in \Sect{sec:train}.

\subsection{Hardware Design}
\label{sec:search:hw}

The hardware designed to exploit the algorithm is shown in \Fig{fig:nsarch}. The search is carried out by a set of PEs, each of which can execute a query independently. The PEs access data from the on-chip SRAM that stores various data structures. The SRAM interfaces with the DRAM through a DMA, as all DRAM accesses are streaming.

\paragraph{SRAM} The SRAM is split into two global buffers and two local buffers. The global tree buffer and query buffer are accessed by all the PEs. Each PE is also equipped with a local result buffer and a local stack buffer private to each query.

The global tree buffer is accessed by the PEs simultaneously. To sustain a high read bandwidth, the tree buffer is heavily banked. Unlike in regular kernels, bank conflicts here could not be avoided by optimizing the data layout in the banks, because the access pattern of the PEs is known only at run time. We will show in \Sect{sec:bank} how to mitigate the performance impact of bank conflicts.


\paragraph{PE Design} The PE design follows the algorithm of how a query traverses the K-d tree to search for its neighbors. As shown in the left blown-up panel in \Fig{fig:nsarch}, a PE is pipelined into five stages, starting from reading the top of the traversal stack (RS) to fetch the next tree node to visit (FN), followed by calculating the distance between the query and the tree node (CD), storing results (SR) when a neighbor is found, and ended with updating the stack (US). The pipeline stalls only when the FN stage meets a bank conflict when reading the global tree buffer.

\paragraph{Hardware Reuse} Due to the uniform traversal-based search in both top- and sub-tree searches, the hardware is reused in both phases. For instance, the PEs are designed with the generic traversal logic that is agnostic to what the search tree is and what the queries are. The US stage is skipped/bypassed in the top-tree search where no backtracking takes place (i.e., no update to the query stack).

The SRAM is also reused between the two phases. Specifically, the PE-local result buffer is re-purposed between storing the sub-tree queues in the top-tree search phase and storing the neighbor results in the sub-tree search phase. The global tree buffer is re-purposed between storing the top-tree and storing the sub-tree. During top-tree search, whenever a result buffer is full all the queries assigned to that queue (thus far) are streamed back to the DRAM.

\subsection{Accuracy and Performance Trade-off}
\label{sec:search:approx}

A key parameter that governs our algorithm is the top-tree height (TTH). TTH must be set to ensure both the top-tree and the sub-trees can be held in the on-chip SRAM. At the same time, TTH also dictates the performance-vs-accuracy trade-off. We explore the implication of TTH in this section.

First, the top-tree height is dictated by the tree buffer size. We require that the entirety of the top-tree or a sub-tree, while is being searched, is completely stored in the tree buffer. This ensures that the PE pipeline does not stall because the required data are off-chip. Thus, the top-tree height $h_t$ must be within the range $[\mathcal{H} + 1 - \log_2 (\mathcal{S}+1),~\log_2 (\mathcal{S}+1)]$ to satisfy the following two inequalities, where $\mathcal{H}$ is the total K-d tree height and $\mathcal{S}$ is the total tree buffer size:
\begin{align}
    2^{h_t} - 1 \leq \mathcal{S} \\
    2^{\mathcal{H} - h_t + 1} - 1 \leq \mathcal{S}
\end{align}


Given that a TTH is within the permissible range, a shorter top-tree increases the neighbor search accuracy at a cost of more computation, and vice versa. This can be explained by a first-order analytical model, where the total number of nodes a query accesses is proportional to the sum of:

\begin{enumerate}[label=(\roman*)]
    \item the number of tree nodes that are visited during the forward traversal, i.e., from the root node of the top-tree to a leaf node of a sub-tree,
    \item the number of nodes that are visited during the sub-tree backtracking.
\end{enumerate}

\begin{figure}[t]
\centering
\begin{minipage}[t]{0.48\columnwidth}
  \centering
  \includegraphics[width=\columnwidth]{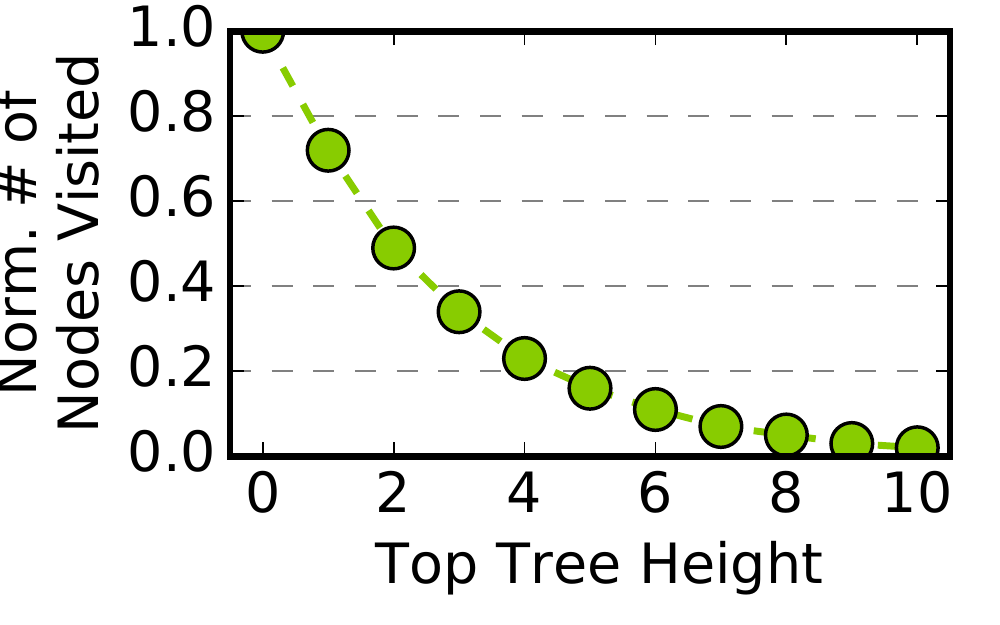}
  \caption{Number of tree nodes visited per query reduces as the top-tree height increases.}
  \label{fig:ht}
\end{minipage}
\hspace{2pt}
\begin{minipage}[t]{0.48\columnwidth}
  \centering
  \includegraphics[width=\columnwidth]{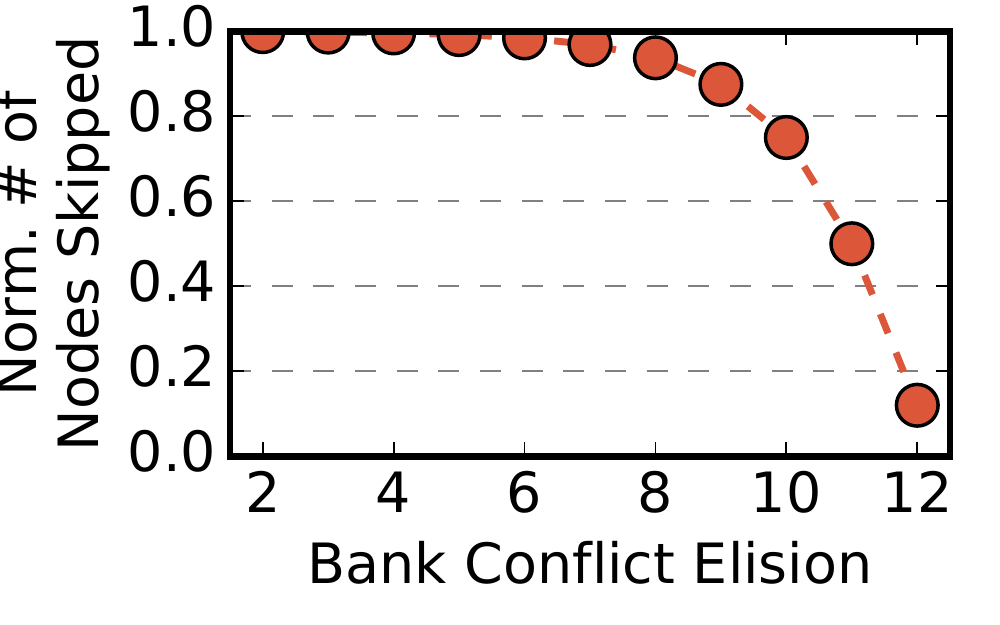}
  \caption{Number of tree nodes skipped per query reduces as the elision height increases.}
  \label{fig:he}
\end{minipage}
\end{figure}

The cost of (i) is constant, as it depends only on the total tree height. The cost of (ii) inversely depends on the TTH: a taller top-tree translates to visiting fewer nodes in the sub-tree backtracking, reducing the cost of (ii) and, by extension, the total cost. \Fig{fig:ht} quantifies how the total number of nodes accessed per query ($y$-axis) varies with the TTH ($x$-axis) using the average statistics of PointNet++(c) on the KITTI dataset. As the TTH increases to 10, only 2\% of the tree nodes are accessed by a query. Visiting fewer nodes improves the search speed but also degrades the accuracy.

An assumption we make, as with Tigris~\cite{xu2019tigris} and QuickNN~\cite{pinkham2020quicknn}, is that a sub-tree can be stored completely on-chip. This is a reasonable assumption: a typical 10 MB point cloud using a 5-level top-tree would result in a sub-tree size of 640 KB, smaller than a typical on-chip buffer size found in mobile SoCs. In case of excessively large point clouds, \proj can in theory recursively split a sub-tree; we do not observe this need in common datasets.

\no{We will show in \Sect{sec:train} that incorporating $h_t$ into the model training process can significantly mitigate the accuracy loss while enabling accuracy-vs-performance trade-off at inference time without retraining.}


\subsection{Efficiency Discussion}
\label{sec:search:eff}

The split-tree algorithm enables completely streaming DRAM accesses. The panel on the right of \Fig{fig:nsarch} shows how the different data structures are laid out in the DRAM and how they are accessed in a streaming fashion. Converting random DRAM accesses to streaming accesses reduces the DRAM energy~\cite{microdrampower, gao2017tetris}, and enables double-buffering, which improves performance because: 1) off-chip data accesses are overlapped with computation, and 2) data needed by the datapath are readily available on-chip without stall.

\no{Specifically, top-tree search involves reading and writing the queries and reading the top-tree nodes, all are streaming. Queries are read in any order they are laid out in the DRAM. Whenever a sub-tree queue is full, all the queries in it are written to the DRAM as a whole, permitting streaming accesses. The sub-tree search streams all the nodes in each sub-tree and all the queries associated with that sub-tree.}

Compared to prior neighbor search algorithms that also enable streaming accesses such as Tigris~\cite{xu2019tigris} and QuickNN~\cite{pinkham2020quicknn}, we reduce both the search load and DRAM traffic. We qualitatively discuss it here, and quantify the gains in \Sect{sec:eval:ns}.

First, Tigris and QuickNN use exhaustive search within the sub-trees, whereas we retain K-d tree search in the sub-trees, thereby reducing the total search load. Retaining K-d tree search in the sub-trees is not an obvious design decision, because it introduces irregular \textit{on-chip} memory accesses in the form of bank conflicts, which prior work aims to avoid at a cost of more search work.

Our strategy is different: we reduce the search work by retaining K-d tree search and mitigate the resulting irregular on-chip memory accesses through inference-training co-design. Specifically, we will show a selective bank conflict elision scheme to significantly reduce bank conflicts (\Sect{sec:bank}), which, when coupled with an approximation-aware training procedure (\Sect{sec:train}), retains the application accuracy.

\no{Second, our algorithm reduces the total amount of DRAM traffic. The reason is that the split-tree structure allows us to search a smaller tree at a time. For instance, with a total tree depth of 10 and a top-tree height of 5, only $\frac{1}{32}$ of the nodes are being searched at once. This dramatic reduction allows us to hold the entire working set (a top/sub-tree being search) on-chip (reducing thrashing) using a small buffer.}

\no{Finally, the top-tree and the sub-tree searches use the same search algorithm (K-d tree traversal). This allows a uniform hardware design: the same hardware can be used in both top-tree search and sub-tree search. This is in stark contrast to prior split-tree algorithms, where the top-tree is searched through tree traversal and the sub-trees are searched using exhaustive search~\cite{xu2019tigris, pinkham2020quicknn}. In that case two completely different hardware modules, one for exhaustive search and one for tree traversal, are instantiated on-chip, leading to a lower area utilization.}

Second, we reduce the amount of DRAM accesses compared to Tigris and QuickNN, both of which load (and reload) a sub-tree from DRAM whenever the corresponding query buffer is full. We instead first stage all the queries to a sub-tree in DRAM and then process them in a batch, thus loading each sub-tree exactly once.

\section{Selective Bank Conflict Elision}
\label{sec:bank}

This section addresses inefficiencies pertaining to on-chip memory accesses. We first describe our main idea of selectively eliding bank conflicts (\Sect{sec:bank:idea}). We then discuss how point cloud algorithms proceed when bank conflicts are elided (\Sect{sec:bank:algo}) and the hardware support (\Sect{sec:bank:hw}). Finally, we identify the key knobs that dictate the accuracy-vs-performance trade-off (\Sect{sec:bank:when}).

\subsection{Main Idea}
\label{sec:bank:idea}

A key requirement of the SRAM design is to feed data required by the PEs without stalling them. Such a requirement is easy to meet in conventional DNNs or other regular kernels, where data access patterns are statically known and thus SRAM data layout can be statically optimized accordingly~\cite{zhou2021characterizing}. The on-chip memory access patterns in point cloud algorithms, however, are only dynamically known, introducing SRAM bank conflicts that are detrimental to overall performance.

Motivated by the error-tolerance nature of neural networks, our idea is to dynamically ignore bank conflicts when appropriate. That is, when multiple memory requests fall in the same bank, instead of serializing the accesses we allow only one request to access the SRAM; other requests return immediately without stalling. While conceptually simple, actually realizing this idea requires answering three questions:

\begin{enumerate}
    \item What data should conflicted accesses return, and how should the algorithm proceed without the correct data?
    \item How to support bank conflict elision in hardware?
    \item When is it appropriate to elide bank conflicts without accuracy drop?
\end{enumerate}

The answers to these questions depend on where a bank conflict takes place in the algorithm, because different memory accesses request data of different significance. Both neighbor search stage and feature computation introduce bank conflicts. In neighbor search, bank conflicts are caused by accessing the tree buffer; all other accesses are regular. In feature computation, aggregating neighbors of a point as inputs to the MLP causes bank conflicts; SRAM accesses incurred during MLP are regular. We now elaborate how the three questions above are addressed in both stages.

\subsection{How Algorithms Proceed with Bank Conflicts Elision}
\label{sec:bank:algo}


\paragraph{Feature Computation} To aggregate neighbors, SRAM accesses are made to retrieve neighbors of a point. Thus, ignoring a conflicted access essentially ignores a point's neighbor, in which case we must fill in the missing neighbors, as the subsequent MLP anticipates an input matrix of a given size (decided at the training time).

To meet the size requirement, we propose to simply reuse the point returned from the request that \textit{is} allowed to access the bank. The intuition is that concurrent accesses, say $A$ and $B$, are guaranteed to be requesting neighbors of the same point $P$~\cite{feng2020mesorasi}. Reusing the returned data from $A$ for $B$ is equivalent to replicating one of $P$'s neighbors. This replication strategy is commonly done in point cloud network design to meet the size requirement in case a neighbor search does not return enough neighbors~\cite{qi2017pointnet, qi2017pointnet++}. Our design essentially performs this replication in hardware, implicitly.

\paragraph{Neighbor Search} The situation is slightly different for neighbor search, where bank conflicts happen when the PEs access the tree nodes during tree traversal. One could use the same replication strategy used in the feature computation stage: if accesses $A_1$ from PE 1 and $A_2$ from PE 2 conflict on the same bank, reuse the data returned from $A_1$ for $A_2$. However, this could lead to side effects such as program crash, redundant computation, and infinite loop. For instance, when the node returned from $A_1$ is in the part of the tree that PE 2 has already visited, pushing $A_1$ onto PE 2's stack leads to an infinite loop or, at least, redundant traversals.

Our design simply ignores the conflicted accesses. Upon a conflict, the FN stage in a PE skips the remaining pipeline stages and reads the next item on the stack. This is denoted by the ``bypass'' signal in the PE shown in \Fig{fig:nsarch}. Algorithmically, this is equivalent to skipping all the nodes beneath the lost node during tree traversal. This strategy omits potential neighbors but guarantees that the traversal terminates.

An optimization that we leave for future work is to check whether the node returned from $A_1$, the request allowed to access the SRAM, is  beneath the node (in the tree) that would have been returned from $A_2$ if the bank conflict were to be observed; if so, using $A_1$ to continue the search in $P_2$ is guaranteed to terminate without side effects. Doing so would skip fewer nodes and potentially increase the accuracy.

\begin{figure}[t]
\centering
\includegraphics[width=\columnwidth]{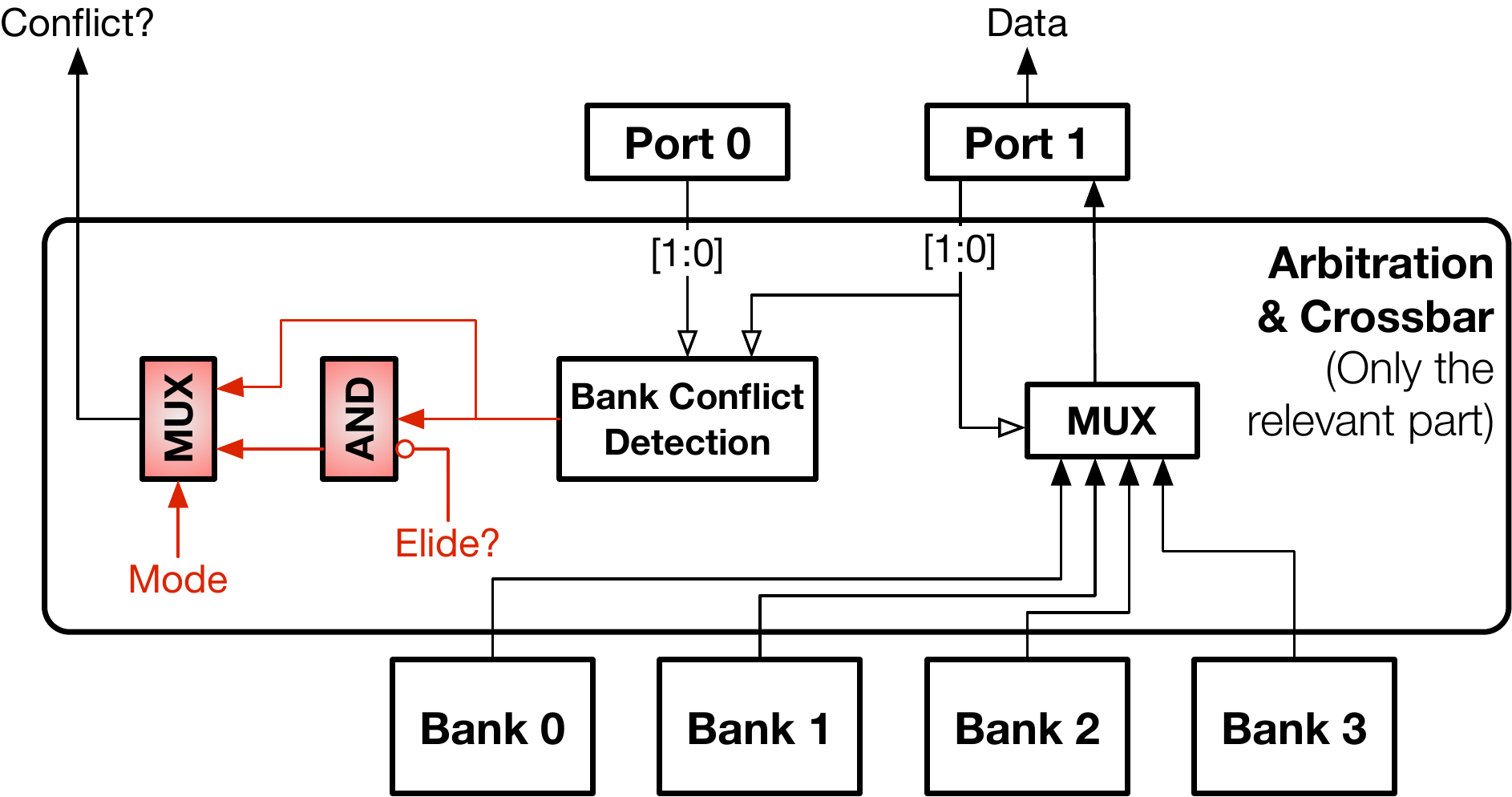}
\caption{Supporting bank conflict elision is trivial in hardware, as many existing hardware structures can be reused. The shaded/colored components are the augmentation, which is required for each SRAM port. Only the relevant part of the hardware is shown for simplicity. The \textit{Mode} signal selects between the neighbor search mode and the feature computation mode. The \texttt{AND} gate lowers the \textit{Conflict} signal when bank conflict elision is enabled in the neighbor search stage.}
\label{fig:bchw}
\end{figure}

\subsection{Hardware Support}
\label{sec:bank:hw}

Eliding bank conflicts is virtually free to implement in hardware by using many existing structures in banked SRAM design. As an example, \Fig{fig:bchw} shows a simple banked SRAM with 2 ports and 4 banks. The key to a banked SRAM is the arbitration and crossbar logic, which detects bank conflicts and routes data from a bank to the right port (a MUX here). For simplicity, we show only the relevant hardware and assume a low-order interleaving, i.e., the two least significant bits in the address select a bank.

Assume both accesses from the two ports fall into Bank 0, and Port 0 is allowed access. In the baseline SRAM, the MUX before Port 1 would select data returned from Bank 0, but this data will be ignored because the bank conflict detection logic would raise the \textit{Conflict} signal, indicating to Port 1 that a bank conflict occurs and the memory request is to be issued again. But, critically, the data returned from Bank 0 is exactly what Port 1 needs in the feature computation stage under bank conflict elision. We simply lower the \textit{Conflict} signal in this case, which is accomplished by \texttt{AND}ing the output of bank conflict detection and the negation of the \textit{Elide} signal, which indicates whether bank conflict elision is enabled.

The \textit{Mode} signal operates a MUX to select between the neighbor search vs. feature computation mode. In neighbor search, the original bank conflict signal is used, except the PE will not re-issue the memory request; instead, the PE simply continues the search with the next item on the stack.

\subsection{When to Elide Bank Conflicts?}
\label{sec:bank:when}

Eliding bank conflicts returns incorrect data to the PEs and, thus, hurts accuracy. We find that eliding bank conflicts in feature computation leads to little to no accuracy loss whereas eliding bank conflicts during neighbor search, without care, has significant accuracy implications (\Sect{sec:eval:train}). This is because in feature computation the data that would have been returned (if bank conflicts were observed) are replaced with the data returned from the conflicting access; in neighbor search, however, eliding bank conflicts directly skips all the computations associated with that node altogether. We thus focus on the neighbor search stage here.

Intuitively, the accuracy loss is smaller when ignoring a memory access made to a lower level tree node, as fewer tree nodes would be skipped later in the traversal. \Fig{fig:he} shows how the percentage of skipped tree nodes ($y$-axis) varies with the tree level below which bank conflicts are elided ($x$-axis). The data are averaged across all the queries of PointNet++(c) on the ModelNet dataset, where the total tree height is 14. When bank-conflicted accesses below level 2 are ignored, almost 100\% of the tree nodes are skipped, which degrades the model accuracy to almost zero (not shown). When the elision level is 12, only 10\% of the tree nodes are skipped.


Skipping more nodes degrades accuracy but increases the search speed. Therefore, a natural knob that controls the trade-off of accuracy-vs-performance is the \textit{elision height} $h_e$, which is defined as the tree level beneath which all conflicted memory accesses are ignored. \Sect{sec:train} will show how incorporating $h_e$ into model training can minimize the accuracy loss while providing the accuracy-vs-performance trade-off without retraining.

\section{Approximation-Aware Network Training}
\label{sec:train}

Our neighbor search algorithm and bank conflict elision, if applied directly on a trained point cloud DNN at inference-time, will decrease the accuracy sharply (\Sect{sec:eval:acc}). This is because the original network is not trained with the various approximation techniques in mind. To mitigate the accuracy drop, we propose a modified network training procedure that mitigates the accuracy loss.\no{ Critically, a co-trained model can adapt to different accuracy-vs-performance trade-offs at inference time without re-training.}




The goal here is to learn a DNN that retains a high accuracy under approximation compared to the baseline network. In particular, we consider two approximation knobs: the top-tree height $h_t$ and the elision height $h_e$.A larger $h_t$ decreases accuracy but increases the performance; conversely, a larger $h_e$ increases the accuracy at a cost of a lower performance.

\begin{figure}[t]
\centering
\includegraphics[width=\columnwidth]{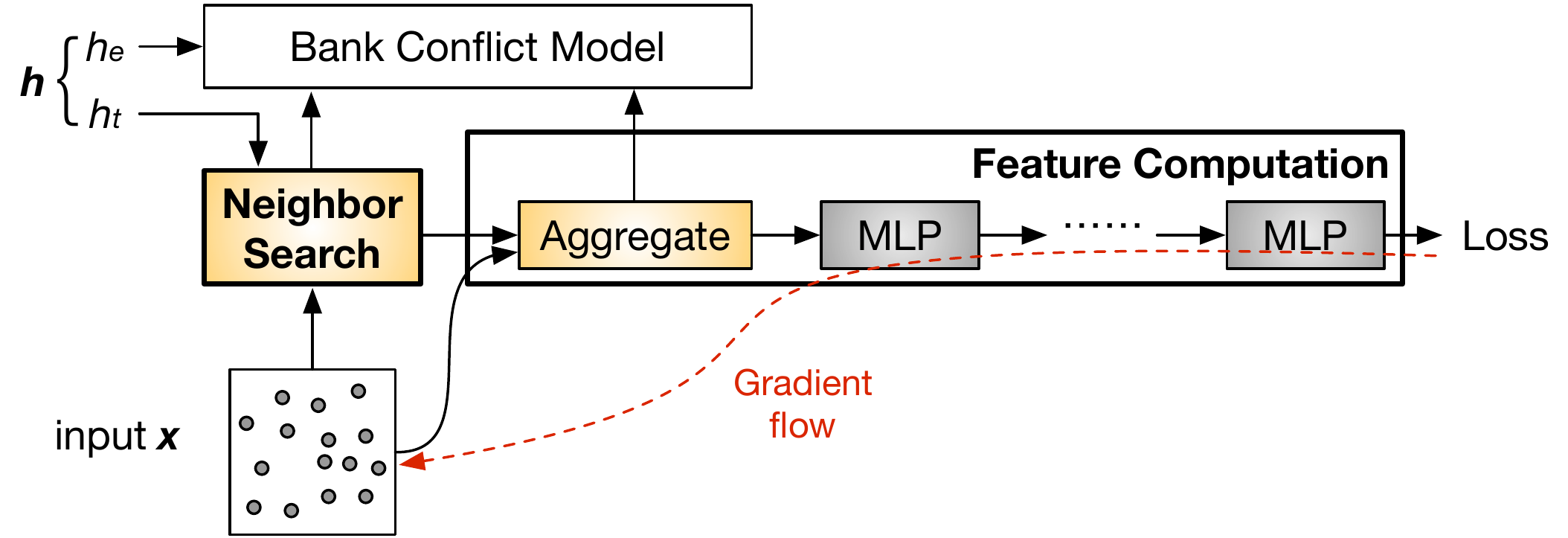}
\caption{Training a point cloud network with approximate neighbor search and bank conflict elision. Note that the training is end-to-end differentiable as in conventional DNN training. The non-differentiable parts, neighbor search and aggregation, do not participate in the gradient flow.}
\label{fig:training}
\end{figure}

A straightforward idea is to integrate $\bh = <h_t, h_e>$ as part of the inference such that the DNN is trained for a particular $\bh$. In essence, this is similar to fine-tuning a compressed model to regain the accuracy, where a network learns to adjust its weights given the approximation introduced by a particular compression setting.

While one could train a dedicated model for each possible $\bh$ and build an ensemble, that would increase the training overhead and deployment complexity. Instead, we propose to learn one model that adapts to different $\bh$. Mathematically, we aim to learn a DNN distribution $f(\cdot, \bh; \theta) \sim F$ such that different DNNs sampled from the distribution $F$ share the same model parameter $\theta$ and provide similar accuracy given an input $\bh$ (along with the input point cloud).

To that end, our training procedure augments the conventional training with one simple extension: conventional training samples input data; our training also randomly samples an $\bh$ \textit{for each input}. During the forward propagation, $\bh$ is used to modulate the neighbor search and bank conflict elision. In this way, the model parameter $\theta$ is trained to accommodate the approximations introduced during the forward inference. The training flow is shown in \Fig{fig:training}.

In order to replay the same inference-time approximation during training, we integrate a hardware simulator for modeling the bank conflict. The bank conflict model is called by both neighbor search and feature computation (the aggregation operation), as \Fig{fig:training} shows. The bank conflict simulator takes in two parameters: 1) $h_e$, which indicates the tree level below which bank conflicts are elided, and 2) the hardware banking configuration (e.g., number of banks, bank size). We find that training with the exact banking configuration on the inference hardware yields higher accuracy, but absent an exact hardware configuration training with a generic banking configuration provides noticeable benefits, too (\Sect{sec:eval:train}).

Finally, note that neighbor search and aggregation do not participate in gradient descent; they simply construct inputs to the MLP layers. Thus, the model is end-to-end differentiable even though neighbor search and aggregation are not.
\section{Experimental Setup}
\label{sec:exp}

\begin{figure}[t]
  \centering
  \includegraphics[width=\columnwidth]{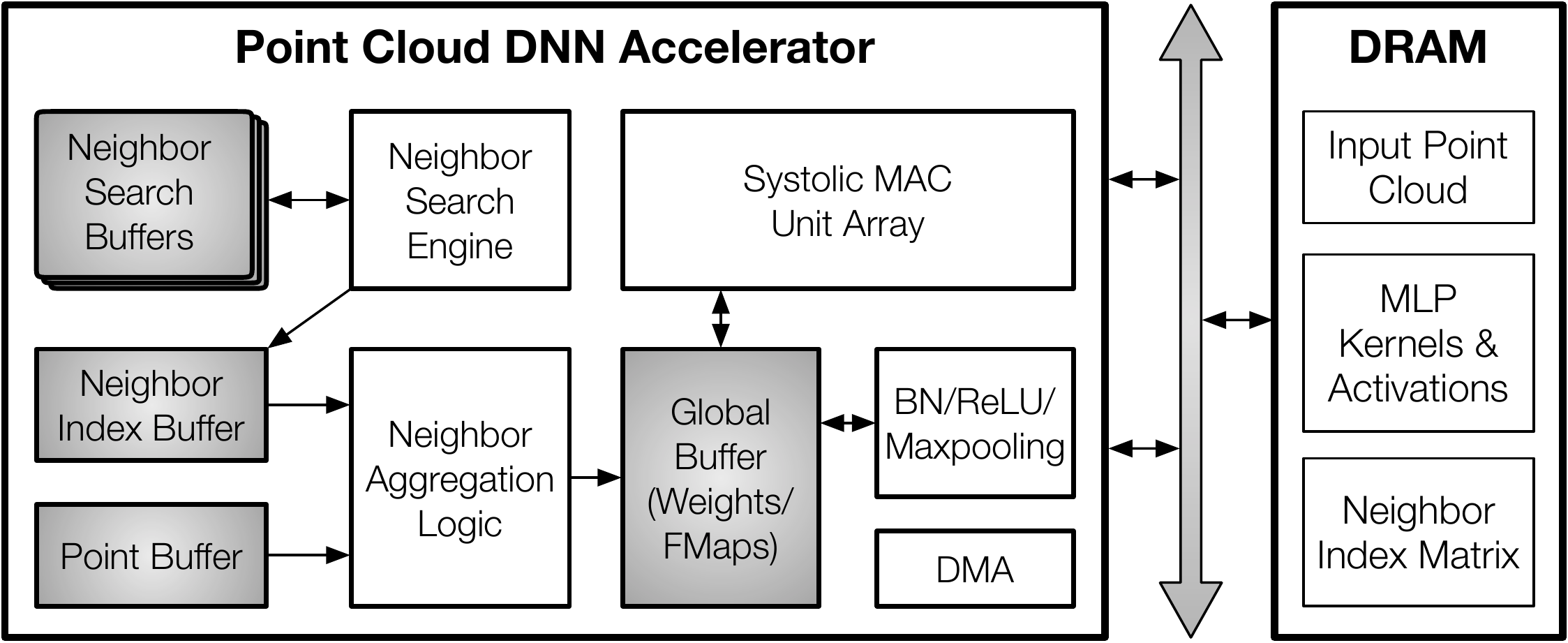}
  \caption{Overall architecture of the point cloud DNN accelerator, which includes three main components: a Neighbor Search Engine, an Aggregation Unit, and a systolic array for executing the MLPs in feature computation. The Neighbor Search Buffers include all the buffers shown in \Fig{fig:nsarch}.}
  \label{fig:arch}
\end{figure}

\paragraph{Architecture Design} \Fig{fig:arch} shows the overall point cloud accelerator, which includes three main components: a neighbor search engine as described in \Sect{sec:search:hw}, a neighbor aggregation unit, which uses the design in Mesorasi~\cite{feng2020mesorasi}, and a DNN accelerator for executing the MLPs. Without losing generality, we assume a systolic-array-based DNN accelerator, which is configured to have a $16 \times 16$ MAC array, where each MAC unit mimics the design of that in the TPU~\cite{tpu}.


The on-chip SRAM is partitioned to serve different purposes. The global buffer serves the weight and activations for the systolic array. It is configured to be 1.5~MB in size. The Point Buffer is a 64 KB 16-banked buffer serving points during aggregation. The Neighbor Index Buffer is sized at 12 KB with a single bank. The Tree buffer and the Query buffer are sized at 6 KB and 3 KB with 4 banks and 1 bank, respectively. These two buffers support selective bank elision as described in \Sect{sec:bank:hw}. The neighbor search engine has 4 PEs, each with a dedicated result buffer and a stack buffer, which are sized at 1.5 KB and 256 B, respectively.



\paragraph{Experimental Methodology} We synthesize, place, and route the datapath of the neighbor search engine, the systolic array, and the aggregation unit using an EDA flow consisting of Synopsys and Cadence tools with the TSMC 16 nm FinFET technology. The SRAMs are generated using the Arm Artisan memory compiler. Power is estimated using Synopsys PrimeTimePX by annotating the switching activity. We then build a cycle-accurate simulator of the architecture with the latency of each component parameterized from the post-synthesis results of the RTL design.

The DRAM is modeled after Micron 16 Gb LPDDR3-1600 (4 channels) according to its datasheet~\cite{micronlpddr3}. The DRAM energy is obtained using Micron System Power Calculators~\cite{microdrampower}. On average, the energy ratio between a random DRAM access and a streaming DRAM access is about 3:1, and the energy ratio between a random DRAM access and an SRAM access is about 25:1, both matching prior work~\cite{gao2017tetris, Yazdanbakhsh2018GAN}.

\begin{table} 
\centering
\caption{Evaluation models.}
\resizebox{.8\columnwidth}{!}{
\renewcommand*{\arraystretch}{1}
\renewcommand*{\tabcolsep}{10pt}
\begin{tabular}{ cccc } 
\toprule[0.15em]
\textbf{\specialcell{Application\\Domains}} & \textbf{\specialcell{Algorithm}} & \textbf{Dataset} \\ 
\midrule[0.05em]
\specialcell{Classification} &  \specialcell{PointNet++ (c)\\DensePoint} & ModelNet40 \\
\midrule[0.05em]
\specialcell{Segmentation} &  
\specialcell{PointNet++ (s)} & ShapeNet \\
\midrule[0.05em]
\specialcell{Detection} &  \specialcell{F-PointNet} & KITTI \\
\bottomrule[0.15em]
\end{tabular}
}
\label{tab:eval_app}
\end{table}

\paragraph{Software Setup} \Tbl{tab:eval_app} lists the four point cloud networks used in the evaluation, which covers common point cloud tasks including classification, segmentation, and detection.
For classification, we evaluate the classic PointNet++(c)~\cite{qi2017pointnet++} and DensePoint~\cite{liu2019densepoint} on the ModelNet40 dataset~\cite{wu20153d}. We use the overall accuracy as accuracy metric. For segmentation, we evaluate PointNet++(s)~\cite{qi2017pointnet++} on the ShapeNet dataset~\cite{shapenet2015}. The metric used in segmentation is the standard Intersection-over-Unit (mIoU) accuracy. For detection, we evaluate F-PointNet~\cite{qi2018frustum} on the KITTI dataset~\cite{geiger2012we} and report the geometric mean of the IoU metric on the car class.

To ensure that the improvements from \proj are not due to the inefficiencies of the network implementation, we use the versions of these models optimized by Feng et al.~\cite{feng2020mesorasi}, which removes redundant MLP computations and on average achieves 1.6$\times$ speedup over the corresponding author-released implementations.

\paragraph{Baseline} We compare against three baselines:

\begin{itemize}
	\item \mode{GPU}: the mobile Pascal GPU on Nvidia's Jetson TX2 development board~\cite{tx2}.
	\item \mode{Tigris+GPU}: this baseline executes the neighbor search on Tigris~\cite{xu2019tigris}, a recent neighbor search accelerator that does not perform approximate eighbor search and selectively bank conflict elision, and executes the feature computation on the mobile Pascal GPU.
	\item \mode{Mesorasi}, a prior point cloud network accelerator~\cite{feng2020mesorasi} that uses Tigris~\cite{xu2019tigris} for neighbor search and executes the feature computation on a dedicated systolic-array without selectively bank conflict elision. The exact same systolic array configuration is used in \proj with the exception that \proj performs selective bank conflict elision.
\end{itemize}



\paragraph{Area Overhead} Our accelerator has a total area of \SI{1.55}{\milli\meter\squared}, in which the \proj-specific portion is almost negligible. The only hardware extension is one that selectively elides the bank conflict (\Fig{fig:bchw}), which requires an additional MUX and an \texttt{AND} gate for each port of the SRAM.

\paragraph{Traininig Overhead} Our approximation-aware training increases the training time by 38\%. The main overhead is to simulate bank conflicts, which currently is a multi-threaded CPU implementation. Using a random $\bh$ does not further increase the training overhead, since we still perform one search per inference. Note that the training overhead is amortized across all subsequent inferences.

\paragraph{Variants} We evaluate two variants of \proj to decouple the contribution of the two optimizations:
\begin{itemize}
    \item \mode{ANS} performs approximate neighbor search but does not elide bank conflicts.
    \item \mode{ANS+BCE} performs approximate neighbor search while also eliding bank conflicts in neighbor search and aggregation.
\end{itemize}
\section{Evaluation}
\label{sec:eval}

We first show that \proj achieves similar accuracy as the baseline (\Sect{sec:eval:acc}) but delivers significant speedups and energy reductions (\Sect{sec:eval:perf}). We then provide a detailed analysis of our training procedure and understand how its effectiveness varies with respect to different algorithmic and hardware configurations (\Sect{sec:eval:train}). We perform a sensitivity study to understand \proj's performance and energy savings vary under different settings (\Sect{sec:eval:sen}). Finally, we provide an quantitative comparison with prior neighbor search accelerators (\Sect{sec:eval:ns}).

\subsection{Accuracy}
\label{sec:eval:acc}

\begin{figure}[t]
  \centering
  \includegraphics[width=\columnwidth]{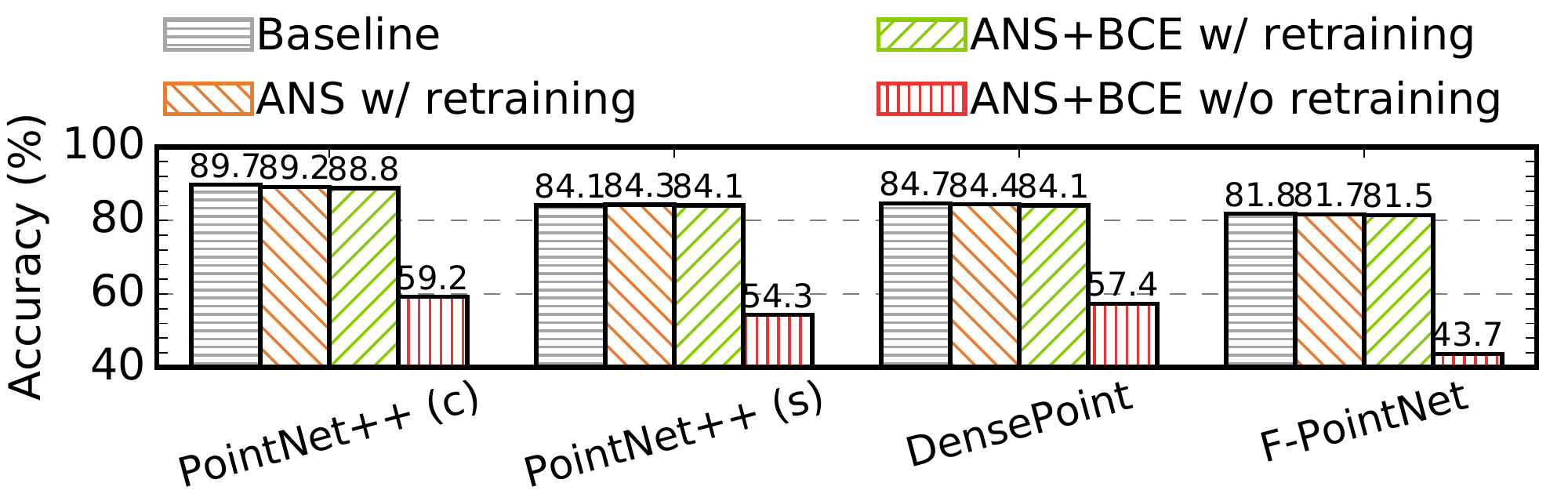}
  \caption{Accuracy comparison between the baseline models, \mode{ANS+BCE} without re-training, \mode{ANS} with re-training under $h_t = 4$, and \mode{ANS+BCE} with re-training under $h_t = 4$ and $h_e=12$.}
  \label{fig:app_acc}
\end{figure}

We find that directly applying \proj optimizations without retraining significantly degrades the model accuracy. Integrating approximation into the training process elevates the accuracy to the baseline level. \Fig{fig:app_acc} compares the model accuracy between four schemes: 1) the baseline models, 2) \mode{ANS+BCE} without re-training, 3) \mode{ANS+BCE} with re-training, and 4) \mode{ANS} with re-training. In this specific case, each re-trained model is trained specifically for the approximate setting where $h_t = 4$ and/or $h_e = 12$.

Directly applying the two optimizations at inference time degrades the accuracy between 27.3\% to 40.5\%, making the models practically useless. Re-training regains the accuracy with an accuracy drop of at most 0.9\% (PointNet++(c)). In PointNet++(s), re-training completely recovers the accuracy loss introduced in approximation. The fact that we can almost completely recover the accuracy loss with \mode{ANS+BCE}, the most aggressive approximation setting, shows the effectiveness of our approximation-aware training. The accuracy of \mode{ANS} alone is slightly higher than that of \mode{ANS+BCE}, as the latter applies two approximations whereas the former applies only one.

\subsection{Performance and Energy}
\label{sec:eval:perf}

\begin{figure}[t]
\centering
\subfloat[Speedup. Higher is better.]{
	\label{fig:hw_speedup}	\includegraphics[width=\columnwidth]{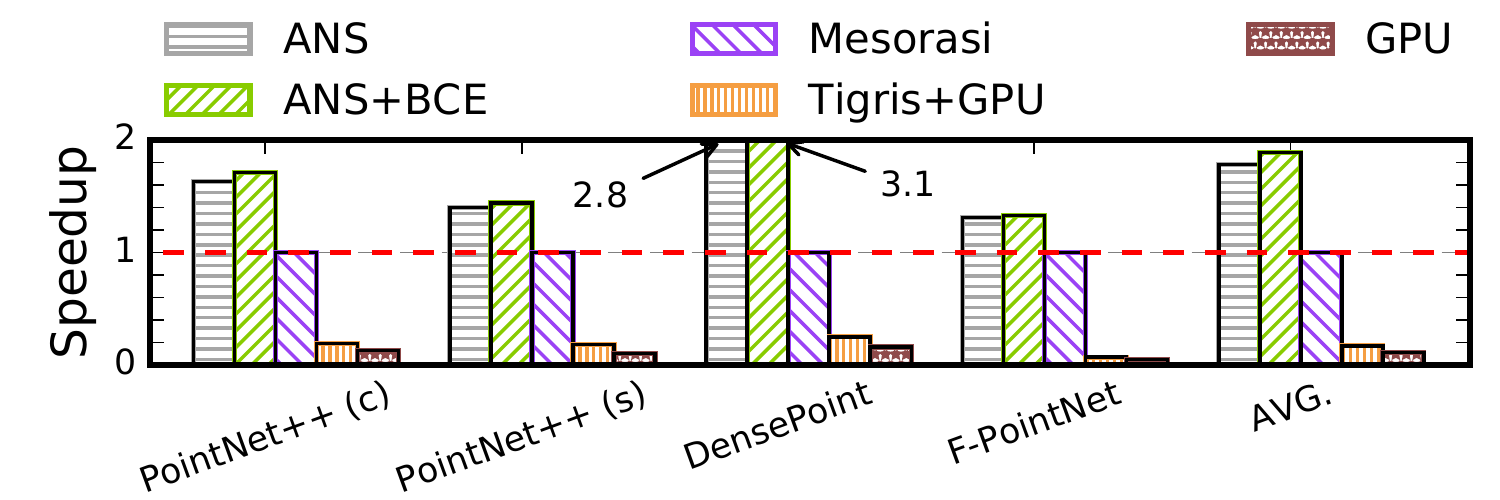} }
\\
\vspace{-5pt}
\subfloat[Normalized energy. Lower is better.]{
	\label{fig:hw_energy}
	\includegraphics[width=\columnwidth]{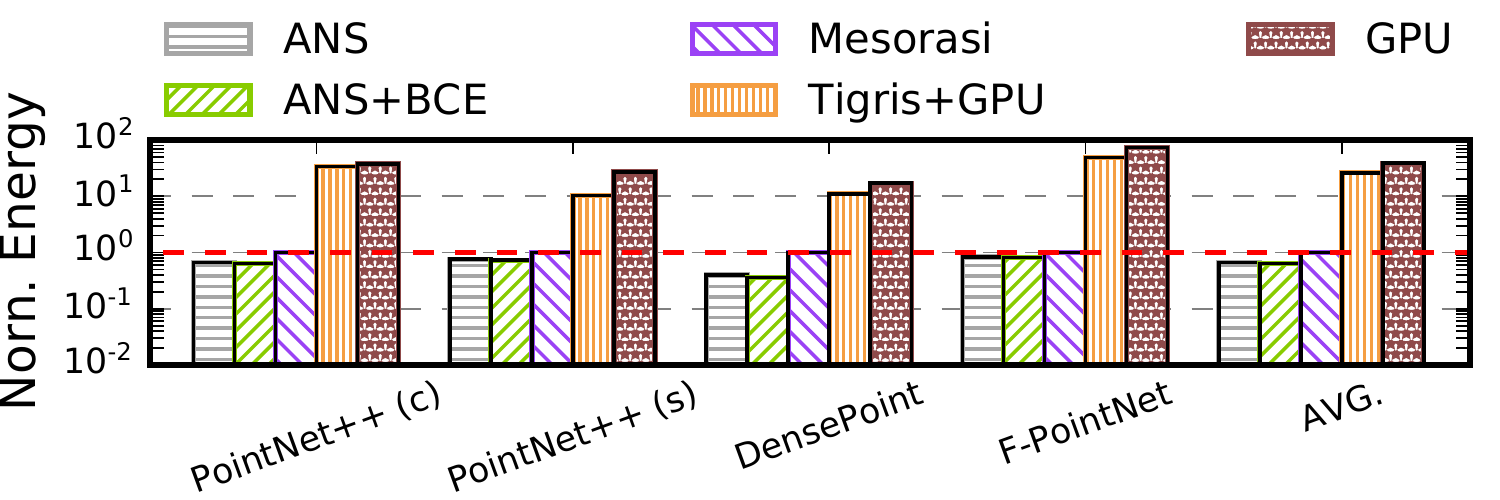} } 
\caption{End-to-end speedup and normalized energy of \mode{ANS} and \mode{ANS+BCE} over the baseline.}
\label{fig:hw_speedup_energy}
\end{figure}

Using the re-trained \mode{ANS} and \mode{ANS+BCE} model shown in \Fig{fig:app_acc}, we compare \proj's performance and energy consumption over the baseline accelerator, shown in \Fig{fig:hw_speedup_energy}.

\paragraph{Speedup} \Fig{fig:hw_speedup} shows the speedup of \mode{ANS} and \mode{ANS+BCE} against the three baselines; all data are normalized to \mode{Mesorasi}. Among the three baselines, \mode{Tigris+GPU} and \mode{GPU} are much slower than \mode{Mesorasi}, because the latter accelerates feature computation on a systolic array.

Overall, \mode{ANS} and \mode{ANS+BCE} achieve a $1.7\times$ and $1.9\times$ speedup, respectively, over \mode{Mesorasi}. Comparing the speed of \mode{ANS+BCE} and \mode{ANS} shows that approximation neighbor search contributes more to the speedup than bank conflict elision. The speedups on DensePoint are the highest ($2.8\times$ and $3.1\times$, respectively) because DensePoint's time is dominated by neighbor search (81\%) whereas neighbor search takes ``only'' about 55\% of the time in other models.

\begin{figure}[t]
  \centering
  \captionsetup[subfigure]{width=0.5\columnwidth}
  \subfloat[\small{Neighbor search.}]
  {
  \includegraphics[width=.48\columnwidth]{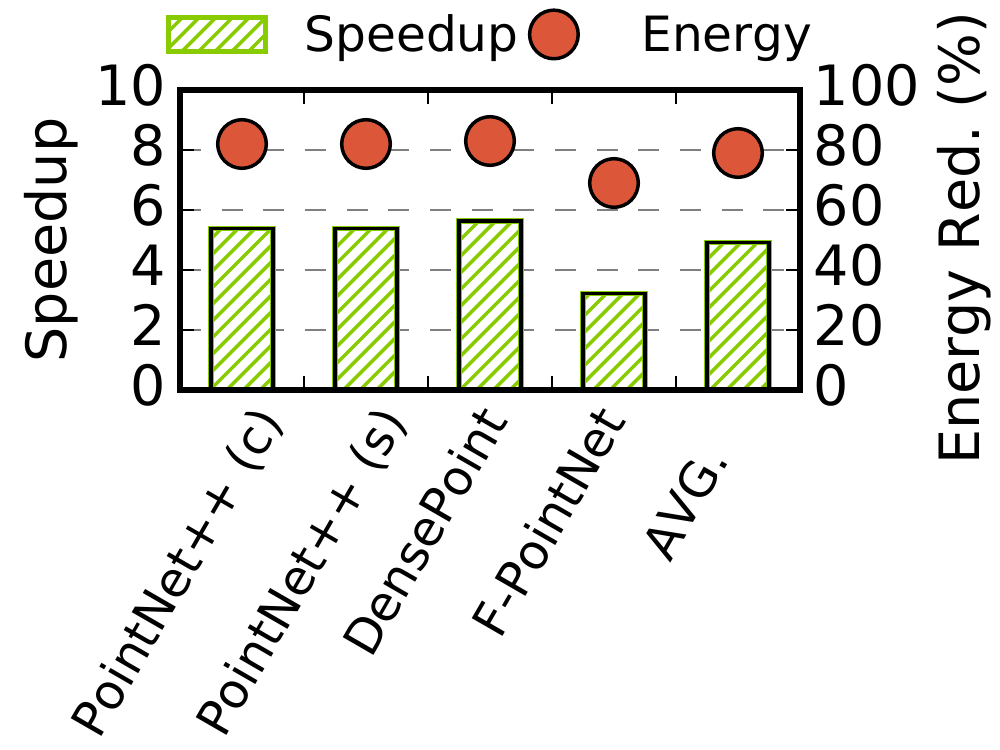}
  \label{fig:ns_speedup}
  }
  \subfloat[\small{Aggregation.}]
  {
  \includegraphics[width=.48\columnwidth]{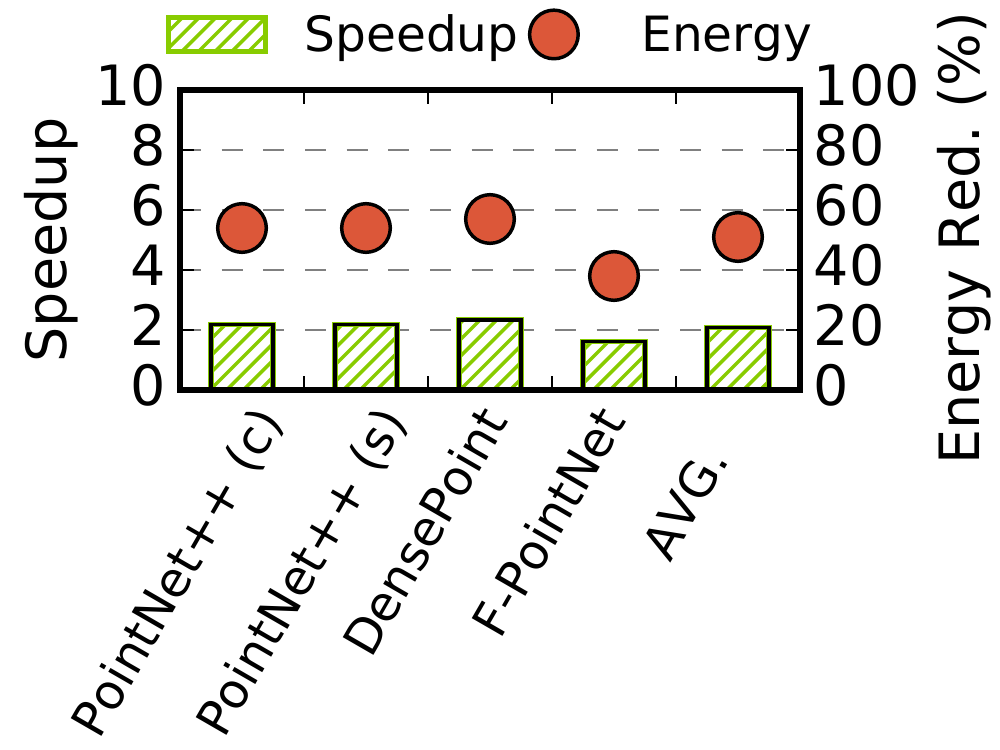}
  \label{fig:aggr_speedup}
  }
  \caption{Speedup and energy savings of \mode{ANS+BCE} on neighbor search and aggregation alone.}
  \label{fig:perf_details}
\end{figure}

To understand the sources of speedup, \Fig{fig:ns_speedup} and \Fig{fig:aggr_speedup} show the speedup of \mode{ANS+BCE} on neighbor search and on the aggregation operation in feature computation, respectively. On average, \mode{ANS+BCE} achieves a $4.9\times$ speedup on neighbor search and a $2.1\times$ speedup on aggregation.

\paragraph{Energy Savings} \Fig{fig:hw_energy} shows the energy consumption of \mode{ANS} and \mode{ANS+BCE} normalized to \mode{Mesorasi}. On average, \mode{ANS} and \mode{ANS+BCE} saves 33\% and 36\% of the total energy, respectively. The energy saving is mainly contributed by approximate neighbor search rather than bank conflict elision, because the former optimizes the DRAM traffic, which contributes more to the energy than the SRAM traffic, which the latter optimizes for. DensePoint, again, has the highest energy saving because it is dominated by neighbor search. As a comparison, \mode{Tigris+GPU} and \mode{GPU} consume $25\times$ and $38\times$ more energy, respectively, compared to \mode{Mesorasi}.

\Fig{fig:ns_speedup} and \Fig{fig:aggr_speedup} on the right $y$-axes show the energy savings on neighbor search and aggregation. DensePoint's savings on these two operations \textit{in isolation} are on par with other networks, confirming that its significant end-to-end savings are primarily attributed to the dominance of neighbor search in its execution time.

\begin{figure}[t]
  \centering
  \includegraphics[width=\columnwidth]{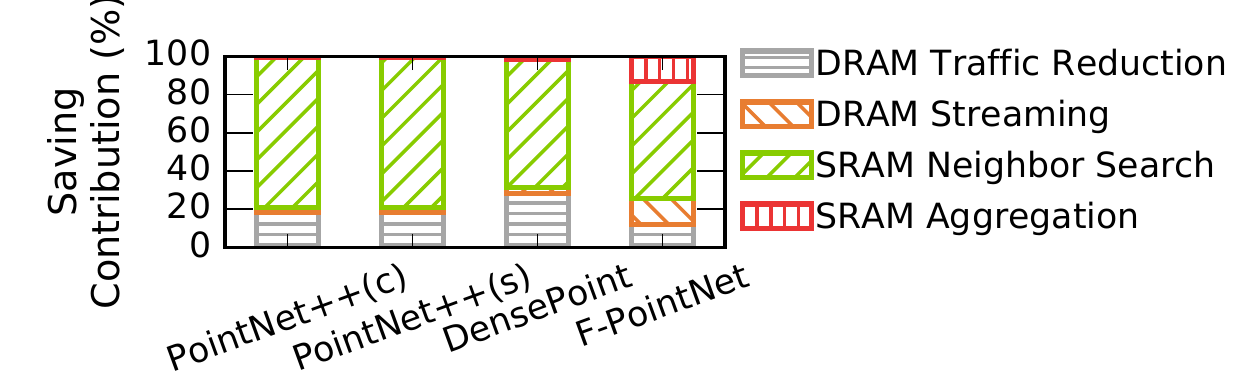}
  \caption{Memory energy saving contribution.}
  \label{fig:saving_dist}
\end{figure}

\begin{figure}[t]
  \centering
  \includegraphics[width=\columnwidth]{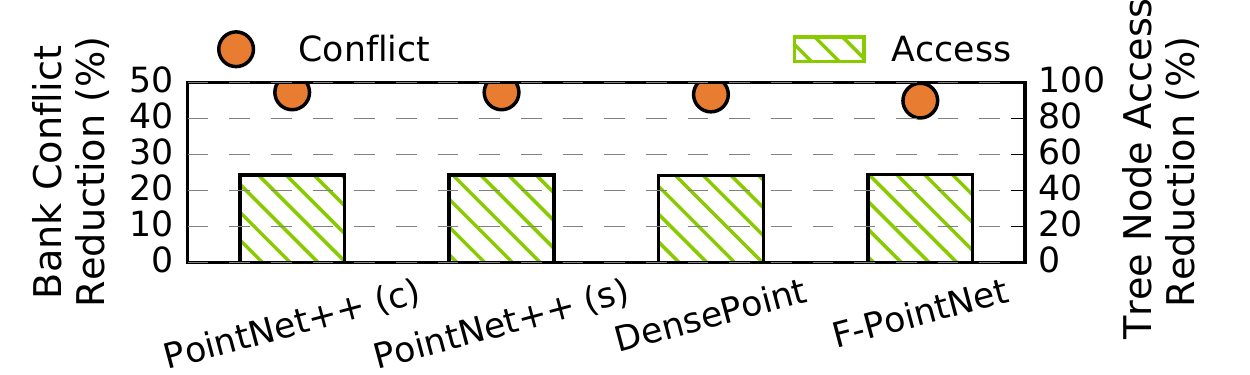}
  \caption{Tree node access saving and bank conflict reduction of \mode{ANS+BCE}.}
  \label{fig:access_and_bank_conflicts}
\end{figure}



\paragraph{Tease Apart Contributions} To understand the sources of energy savings, \Fig{fig:saving_dist} decouples the memory energy savings into four components: converting random DRAM accesses to streaming accesses, DRAM traffic reduction, SRAM traffic reduction in neighbor search, and SRAM traffic reduction from aggregation. The former two are due to the new neighbor search algorithm, and the latter two are due to the selective bank conflict elision optimization.

The main energy saving contributor is the SRAM traffic reduction in neighbor search, which frequently accesses the Tree Buffer. While the DRAM savings are relatively smaller, we expect the DRAM savings will become more significant in the future as the point clouds grow in size.

We quantify the impact of selective bank conflict elision (BCE) in \Fig{fig:access_and_bank_conflicts}, where we show the reduction in bank conflicts (left $y$-axis) and, as a result, the reduction in the number of tree nodes visited (right $y$-axis). The results are obtained by comparing \mode{ANS+BCE} with \mode{ANS}. Overall, BCE avoids over 45\% of bank conflicts and reduces 50\% of tree node accesses in neighbor search. This result explains the $1.9\times$ speedup over \mode{Mesorasi} by \mbox{\mode{ANS+BCE}}.

%
%
%

\subsection{Understanding the Training Procedure}
\label{sec:eval:train}

\begin{figure}[t]
\centering
\begin{minipage}[t]{0.48\columnwidth}
  \centering
  \includegraphics[width=\columnwidth]{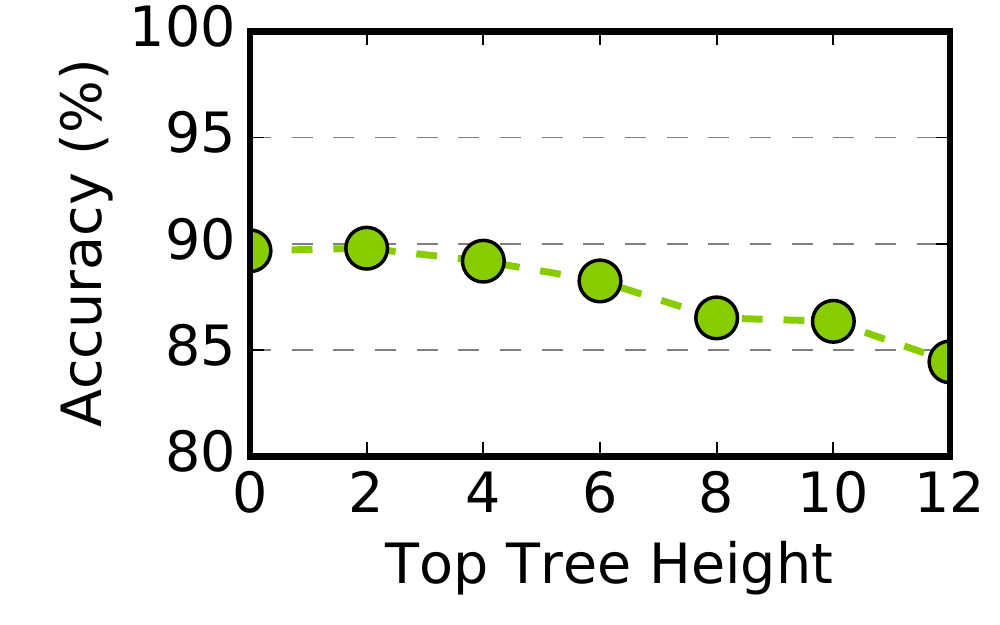}
  \caption{Accuracy of dedicated PointNet++(c) models under different top-tree heights ($h_t$).}
  \label{fig:top_tree_height}
\end{minipage}
\hspace{2pt}
\begin{minipage}[t]{0.48\columnwidth}
  \centering
  \includegraphics[width=\columnwidth]{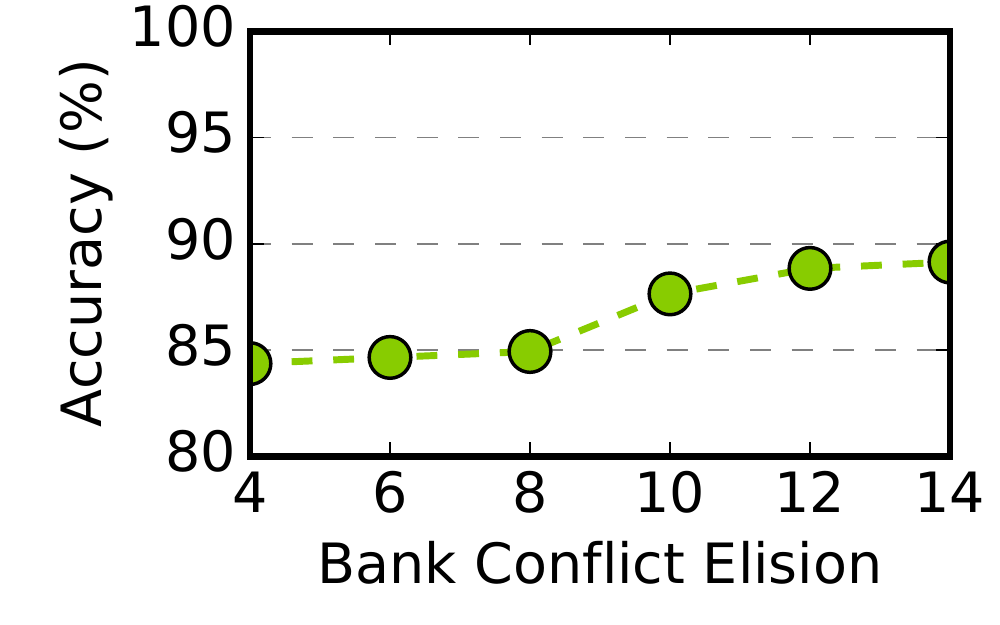}
  \caption{Accuracy of dedicated PointNet++(c) models under different elision heights ($h_e$).}
  \label{fig:ibc_height}
\end{minipage}
\end{figure}

We use PointNet++(c) as a representative model to drive the analyses in this section. The conclusions generally hold.

\paragraph{Dedicated Models} We first evaluate the accuracy of models trained with dedicated approximation settings.

\Fig{fig:top_tree_height} shows the accuracy of PointNet++(c) trained under different top-tree heights ($h_t$) and then inferenced under the same $h_t$. The setting $h_t$ being 0 is the baseline model with exact search. As the $h_t$ increases, the accuracy decreases. This is because a larger $h_t$ reduces the search space and, thus, it is less likely to find the exact neighbors for each query. The accuracy is acceptable initially, dropping from 89.6\% to 88.8\% as $h_t$ increases from 0 to 4. Beyond 4, the accuracy drop becomes more significant. As the top-tree height reaches 12, the accuracy is only 84.4\%. As we will show later, however, a higher $h_t$ leads to a higher speedup, providing a large trade-off space.

\Fig{fig:ibc_height} performs a similar study while varying the elision height $h_e$. Each marker in the figure represents a dedicated \mode{ANS+BCE} model trained with different $h_e$ ranging from 4 to 14; $h_t$ in this example is fixed at 4. As $h_e$ increases, the accuracy increases. This is because a higher elision height skips fewer tree nodes during tree traversal, leading to a better search result. At a $h_e$ of 12, the accuracy loss is only 0.8\%. The accuracy loss is over 5\% when $h_e$ reduces to 4, essentially ignoring almost all nodes in the sub-tree.

\begin{figure}[t]
\centering
\begin{minipage}[t]{0.48\columnwidth}
  \centering
  \includegraphics[width=\columnwidth]{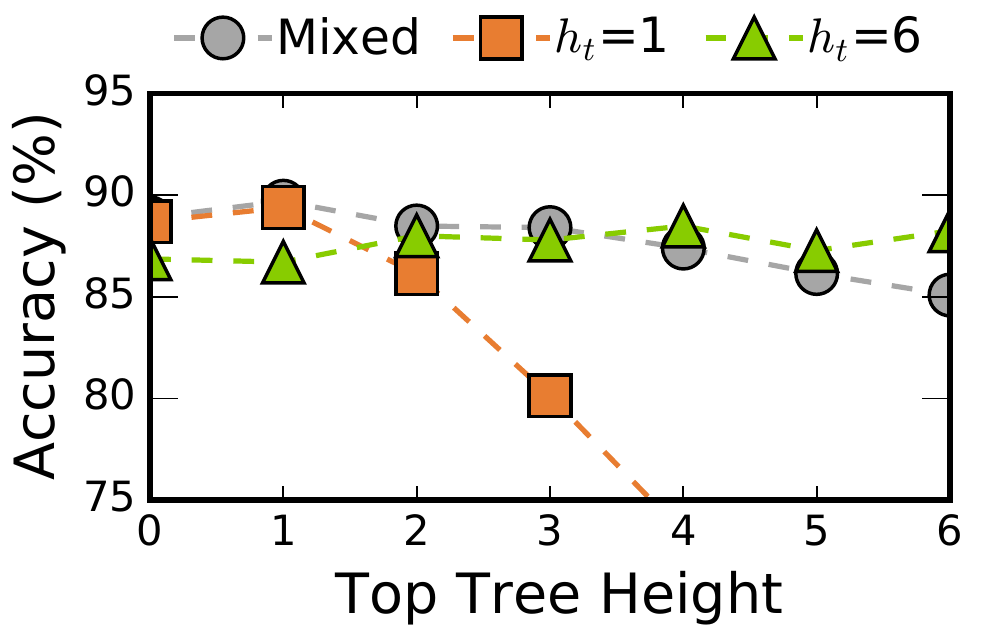}
  \caption{Accuracy comparison of different training schemes.}
  \label{fig:train_comp}
\end{minipage}
\hspace{2pt}
\begin{minipage}[t]{0.48\columnwidth}
  \centering
  \includegraphics[width=\columnwidth]{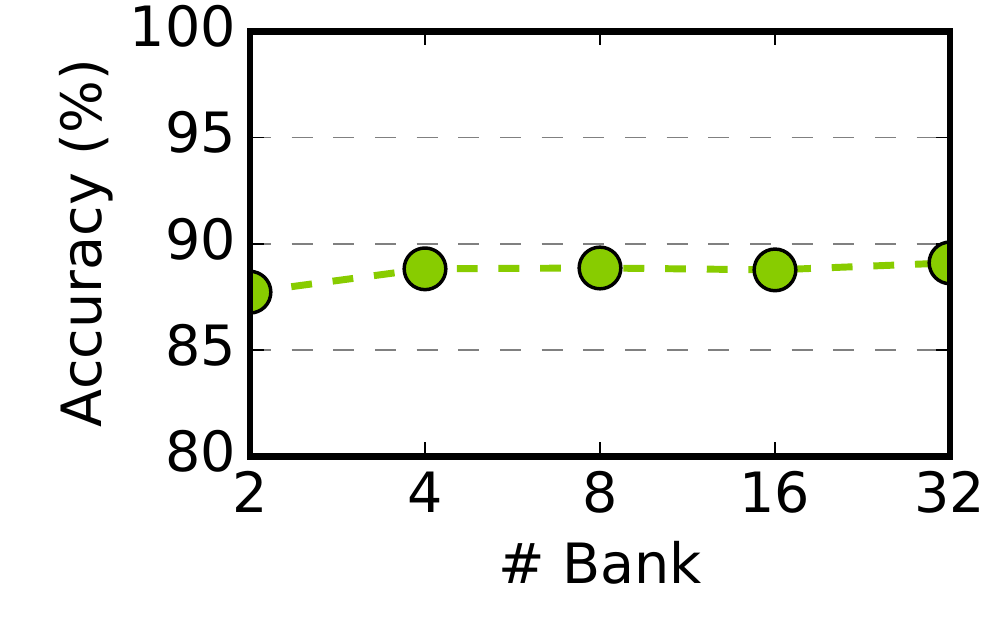}
  \caption{Sensitivity of training accuracy to bank conflict configuration.}
  \label{fig:bank_acc}
\end{minipage}
\end{figure}

\paragraph{Mixed Training} We now evaluate how a model trained by sampling approximation settings adapts to different approximation levels at inference time. \Fig{fig:train_comp} compares three schemes: 1) a model trained with $h_t=1$, 2) a model trained with $h_e=6$, and 3) a model trained by random sampling $h_t$ between 1 and 6 for each input (``Mixed'' in the figure). We show their accuracy under different inference-time $h_t$.

When a dedicated model is trained with $h_t=1$, the accuracy significantly drops when the inference-time $h_t$ is greater than 1. This is not surprising: a model trained with little approximation in mind does not perform well when inference performs aggressive approximation. When a dedicated model is trained with $h_t=6$, however, it performs reasonably well across different $h_t$ at inference-time, even for $h_t$ settings that are not seen in the training time.

The mixed model consistently provides higher or similar accuracy compare the dedicated $h_t=1$ model. Compared to the dedicated $h_t=6$ model, the mixed model is significantly better when higher accuracy is required (i.e., $h_t \leq 3$). The accuracy is only noticeably worse than the dedicated $h_t=6$ model when the inference-time $h_t$ is 6, which is what the dedicated $h_t=6$ model is trained to do well on. The mixed model is favorable when accuracy requirement is high, which is arguably more important than the low-accuracy regime.


\paragraph{Bank Conflict Simulation} In order to integrate bank conflict elision into training, we simulate the bank conflicts in the forward propagation process during training. However, at training time the exact banking configuration of the target hardware might be unknown. \Fig{fig:bank_acc} show the accuracy of training a model assuming 4 banks in the SRAM while inferencing under different numbers of banks. The accuracy beyond 8 is largely stable; the accuracy has about 2\% drop when inferencing on a 2-banked SRAM. 

\paragraph{BCE in Aggregation vs. Neighbor Search} We perform bank conflict elision in both neighbor search and in feature aggregation. We find that the overall accuracy is insensitive to bank conflict elision in aggregation even \textit{without} re-training. Across all networks, applying bank conflict elision in aggregation alone (while turning off other approximations) results in at most 0.3\% accuracy loss. In contrast, accuracy typically drops by double digits if bank conflict elision is applied in neighbor search without re-training. As discussed in \Sect{sec:bank:when}, this is because in the latter case eliding bank conflicts completely skips subsequent search operations.

\subsection{Sensitivity Study}
\label{sec:eval:sen}

\begin{figure}[t]
  \centering
  \captionsetup[subfigure]{width=0.5\columnwidth}
  \subfloat[\small{Speedup sensitivity.}]
  {
  \includegraphics[width=.48\columnwidth]{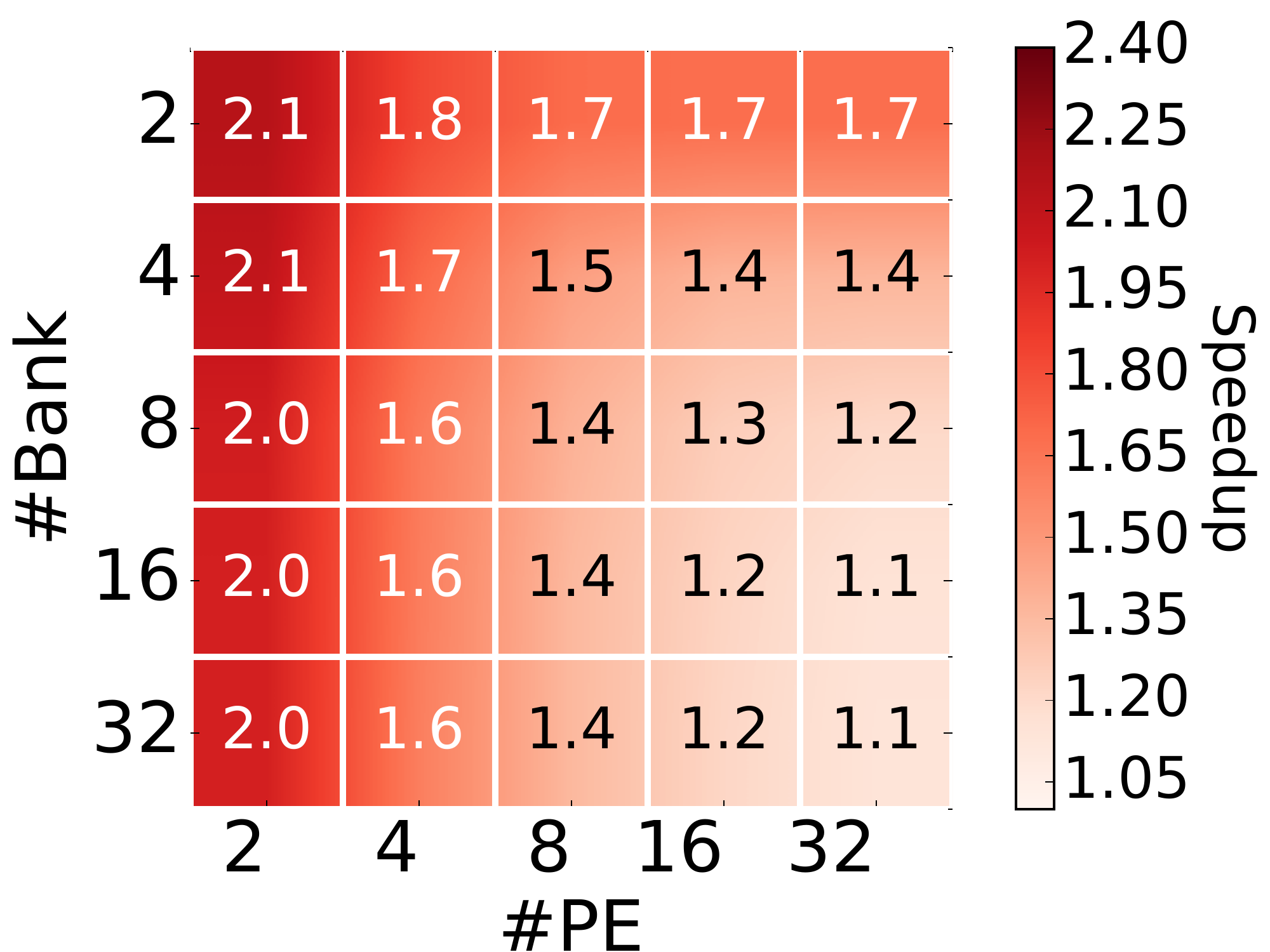}
  \label{fig:profile_speedup}
  }
  \subfloat[\small{Energy sensitivity.}]
  {
  \includegraphics[width=.48\columnwidth]{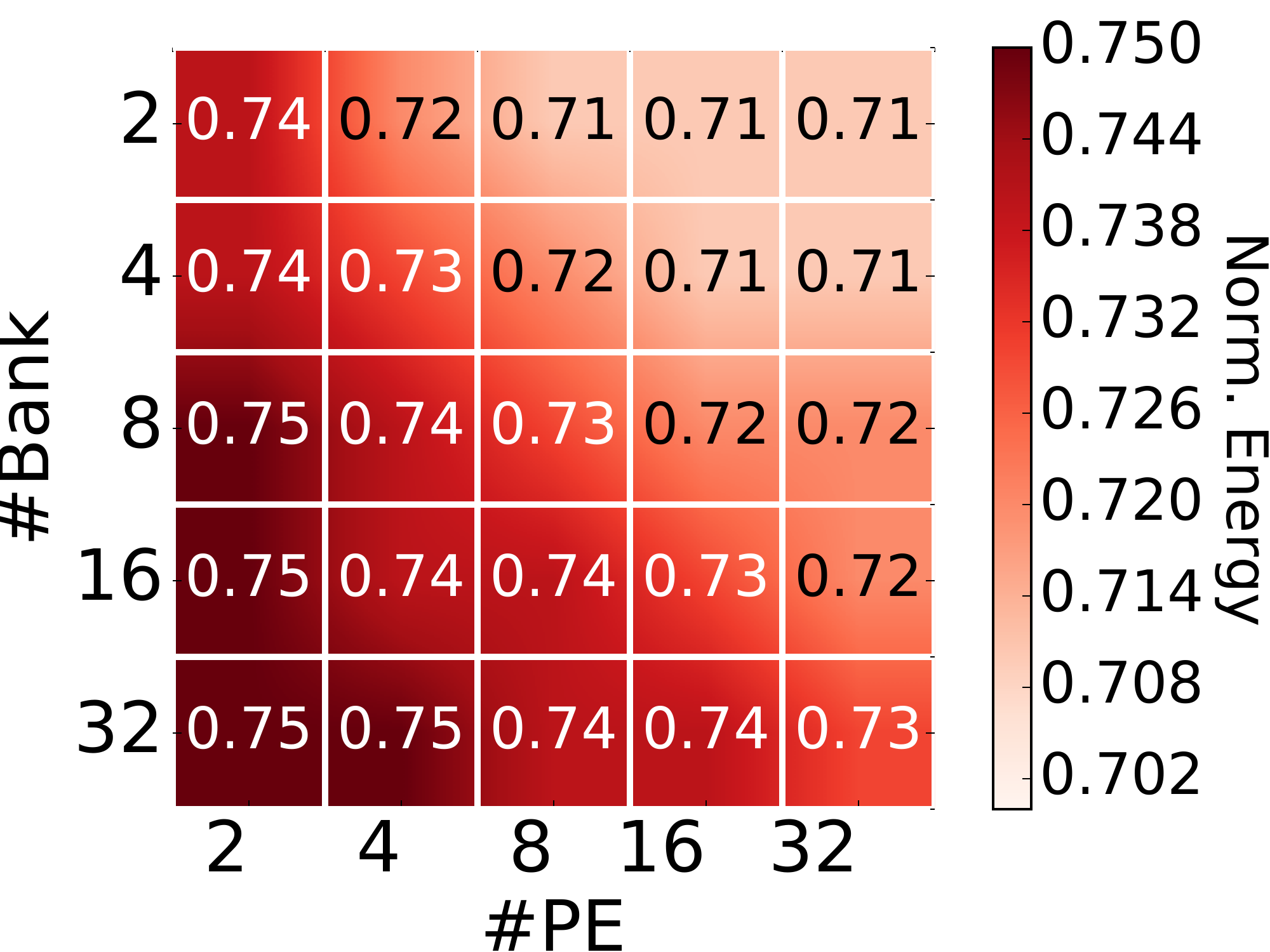}
  \label{fig:profile_energy}
  }
  \caption{Sensitivity of speedup and (normalized) energy to hardware configuration (PE and bank counts) on PointNet++(c).}
  \label{fig:sen}
\end{figure}

\paragraph{Hardware Configuration} \Fig{fig:profile_speedup} and \Fig{fig:profile_energy} show how \proj's speedup and energy vary, respectively, as the numbers of PEs and the number of Tree Buffer banks vary. The energy is normalized to the corresponding baseline.

Naturally, the speedup is higher on less-capable baselines and diminishes on more capable baselines (e.g. 32 PEs and 32 banks), because performance optimizations are less important when the hardware is faster to begin with. Note, however, that a 16-bank memory introduces a large cross-bar overhead and is generally impractical for mobile-grade accelerators~\cite{agarwal2009garnet, zhou2021characterizing}.

The significant energy saving is consistent across hardware configurations. Even with a 32 PE 32 bank configuration, \proj still saves about 27\% energy on PointNet++(c). This is because the energy is roughly proportional to the amount of work done. Changing the hardware configuration does not affect the bulk of the work needed to be done.

\begin{figure}[t]
  \centering
  \captionsetup[subfigure]{width=0.5\columnwidth}
  \subfloat[\small{Speedup-vs-accuracy.}]
  {
  \includegraphics[width=.48\columnwidth]{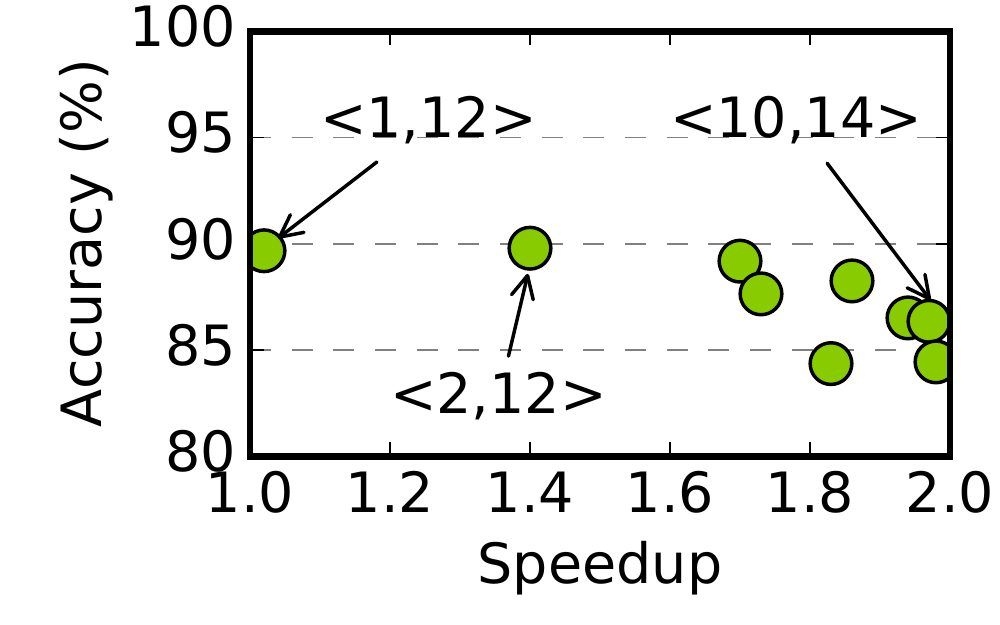}
  \label{fig:speedup_scatter}
  }
  \subfloat[\small{Energy-vs-accuracy.}]
  {
  \includegraphics[width=.48\columnwidth]{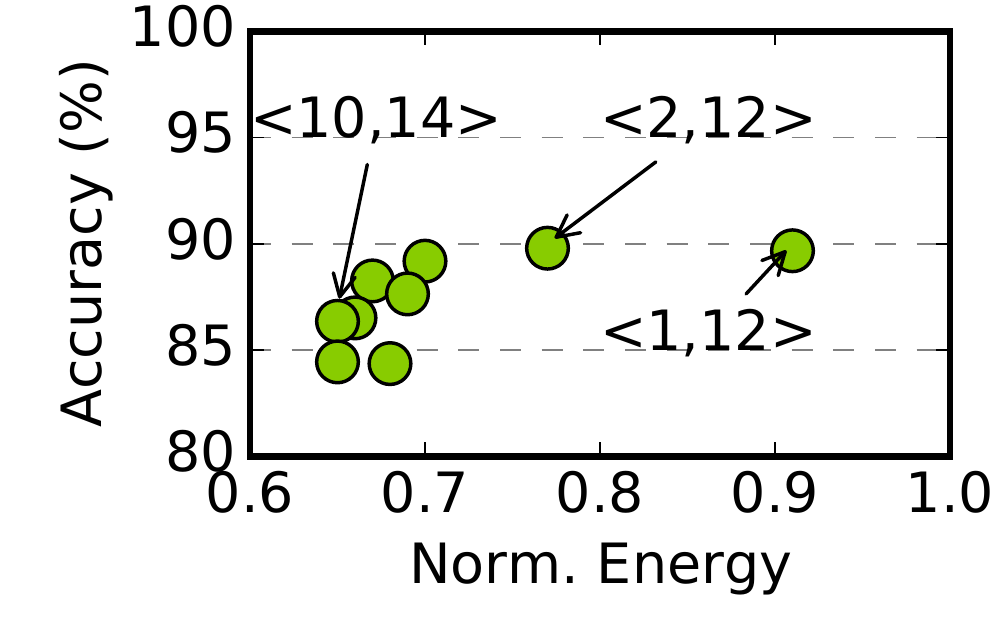}
  \label{fig:energy_scatter}
  }
  \caption{Accuracy vs. performance vs. energy trade-off on PointNet++(c) under different <$h_t,  h_e$> combinations.}
  \label{fig:scatter}
\end{figure}

\paragraph{Approximation Degrees} \Fig{fig:speedup_scatter} and \Fig{fig:energy_scatter} show the accuracy-vs-speedup and accuracy-vs-energy trade-offs, respectively, with different $h_t$ and $h_e$ combinations, which dictate different approximation strengths. The data are reported from PointNet++(c), but the trend generally holds. Overall, varying $h_t$ from 0 to 12 and $h_e$ from 4 to 14 provide a trade-off space of about 5\% accuracy range, 2.0 $\times$ performance range, and 1.5 $\times$ energy range.

\subsection{Comparison with Prior Neighbor Search Accelerators}
\label{sec:eval:ns}

\begin{figure}[t]
  \centering
  \captionsetup[subfigure]{width=0.48\columnwidth}
  \subfloat[\small{Reduction in tree nodes visited from Tigris~\cite{xu2019tigris}.}]
  {
  \includegraphics[width=.48\columnwidth]{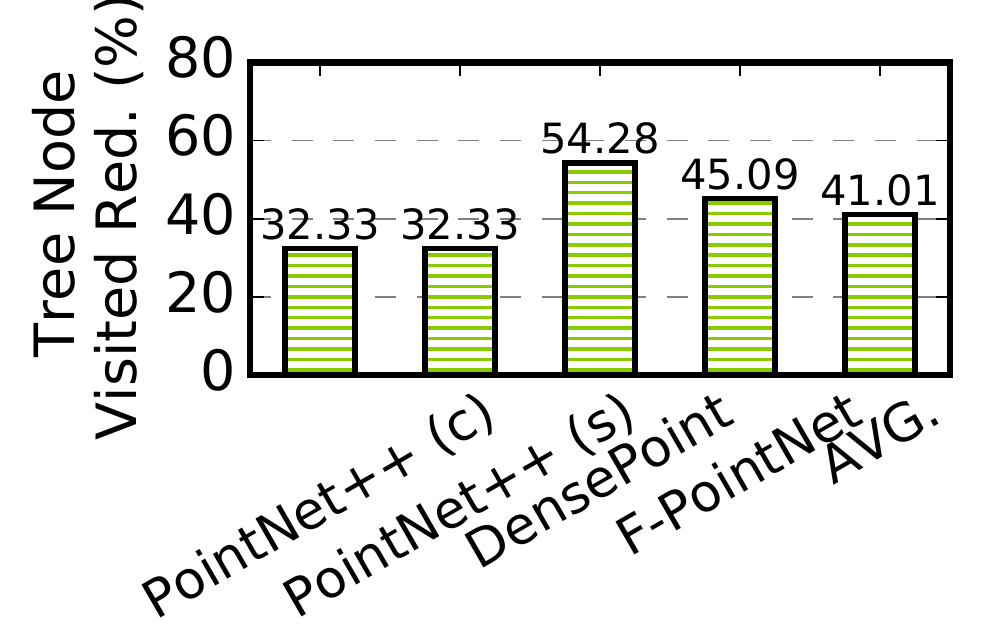}
  \label{fig:compare_tigris}
  }
  \subfloat[\small{DRAM access (in Bytes) reduction from QuickNN~\cite{pinkham2020quicknn}.}]
  {
  \includegraphics[width=.48\columnwidth]{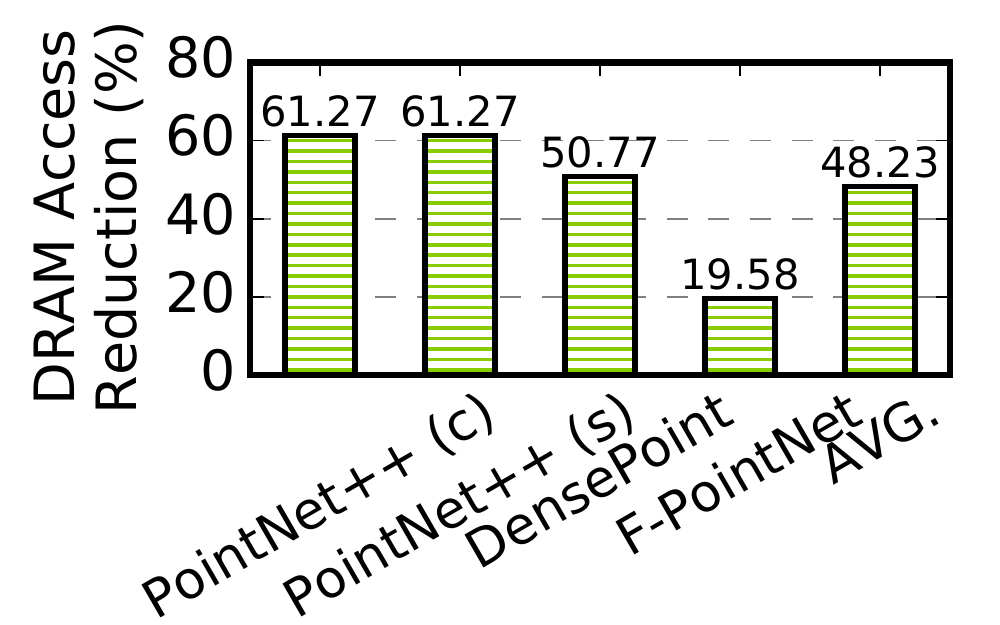}
  \label{fig:compare_quicknn}
  }
  \caption{Comparison with prior neighbor search accelerators Tigris and QuickNN.}
  \label{fig:cmp_tigris_quicknn}
\end{figure}

QuickNN~\cite{pinkham2020quicknn} and Tigris~\cite{xu2019tigris} are two recent neighbor search accelerators that both use a split-tree data structure. As discussed in \Sect{sec:search:eff}, \proj reduces both the search load and DRAM traffic. \Fig{fig:compare_tigris} shows that the K-d tree-based search reduces the total number of tree nodes visited by 41\% compared to exhaustive search. This explains the one order of magnitude performance improvement over the Tigris-based accelerator shown in \Sect{sec:eval:perf}.

QuickNN, similar to \proj, also presents streaming DRAM accesses --- at the expense of redundant DRAM accesses, since each sub-tree is potentially loaded onto the accelerator multiple times. Comparing to a QuickNN implementation with the same PE configuration, \Fig{fig:compare_quicknn} shows that \proj reduces the total DRAM accesses by \mbox{48\%}.

Finally, we target DNN-based algorithms and, thus, can mitigate the potential accuracy loss through end-to-end network training, which is not available to QuickNN and Tigris; both target a non-DNN algorithm (point cloud registration).


\section{Related Work}
\label{sec:related}


\paragraph{Deep Learning for Point Clouds} Point cloud algorithms are increasingly moving toward DNNs, which has spurred recent interests in accelerating point cloud networks~\cite{lin2021pointacc, feng2020mesorasi, hyun2021characterization}. Point cloud DNNs mainly come in two forms: one that operates on raw points~\cite{qi2017pointnet++, qi2017pointnet, wang2019dynamic, zhang2019linked, liu2019densepoint}, and the other that first voxelizes points and operates on voxels, which are grid-aligned points~\cite{choy20194d, graham20183d}. The former requires explicitly neighbor search whereas the latter accesses neighbors through simple indexing. It is unclear whether future point cloud algorithms will definitively favor one form over the other. \proj focuses on optimizing point-based algorithms, whose flexibility and compact data representations are shown to be critical in many application domains~\cite{guo2020deep}, such as object detection, localization (SLAM), segmentation, and classification.

PointAcc~\cite{lin2021pointacc}, Point-X~\cite{zhang2021point}, and Mesorasi~\cite{feng2020mesorasi} are all recent point cloud accelerators. They are fundamentally orthogonal to our work in that they focus on accelerating the feature computation in point cloud DNNs. For instance, Point-X and Mesorasi exploit the spatial locality and computation redundancy, respectively. All three use brute-force neighbor search and, thus, can directly benefit from the optimizations (approximate neighbor search and selective bank conflict elision) proposed in this paper. We show 1.9 $\times$ speedup and 36\% energy reduction over Mesorasi in \mbox{\Sect{sec:eval:perf}}.

\paragraph{Neighbor Search} This paper targets neighbor search in low-dimensional space (2/3D), which is a fundamental building block in many computational science and engineering fields, where physical objects naturally lie in 2/3D space, such as computational fluid dynamics~\cite{ihmsen2011parallel}, computer graphics~\cite{yifan2019differentiable}, and vision~\cite{lu20213d, xu2019tigris}. Prior work has explored both algorithmic and hardware solutions to accelerate neighbor search~\cite{aly2011distributed, Qiu2009GPU, Gieseke2014Buffer, Heinzle2008A, Kuhara2013An, Winterstein2013FPGA, pinkham2020quicknn, xu2019tigris}, many of which are approximate in their nature~\cite{arya1998optimal, miclet1983approximative, greenspan2003approximate, ma2002low, purcell2005photon, Heinzle2008A}. We provide a quantitative comparison with QuickNN~\cite{pinkham2020quicknn} and Tigris~\cite{xu2019tigris}, two most relevant accelerators in \Sect{sec:eval:ns}.

\paragraph{Optimizing Irregular Memory Accesses} Recent work has made significant strides in domain-agnostic prefetching for irregular applications~\cite{ainsworth2018event, talati2021prodigy, naithani2021vector}. Our split-tree structure can be seen as an application-specific prefetcher and achieves ``perfect prefetching'' in that 1) off-chip data accesses are overlapped with computation, 2) data needed by the accelerator are readily available on-chip without stalls, and 3) no redundant DRAM accesses are needed.

Our split-tree structure also serves as an irregular tiling strategy, akin to propagation blocking for graph algorithms~\cite{beamer2017reducing}, but the decision as to which partition (sub-tree) a point is stored is based on the geometric position of a point.

\no{Due to the curse of dimensionality, however, neighbor searches in high dimensional space require fundamentally different algorithms than those in low-dimensional space~\cite{anna, beyer1999nearest, weber1998quantitative, walters2010comparative}, and are out of this paper's scope.}


\paragraph{Approximation Techniques} Our approximation techniques exploit the inexact nature of DNNs. Selective bank conflict elision can be seen as a form of value approximation, bearing similarity to such approximation in general-purpose processors~\cite{san2014load, miguel2015doppelganger, san2016bunker, rengasamy2015exploiting, wong2016approximating}. However, different from prior systems where the accuracy control is empirical, we integrate approximation into the training process; this allows us to provide statistical accuracy guarantees.

%

\section{Conclusion}
\label{sec:conc}

The mismatch between 3D perception algorithms and today's hardware designed and optimized for 2D perception will only increase in the future, where 3D perception applications are expected to be much more compute- and memory-intensive while at the same time being deployed in more resource-constrained platforms such as micro aerial vehicles.

The mismatch between 3D perception algorithms and today's hardware designed and optimized for 2D perception will only increase in the future. \proj demonstrates an algorithm-hardware collaborative approach toward taming the irregularities in point cloud algorithms. The key idea behind \proj is to intentionally introduce approximations at both the algorithm and the hardware level to reduce memory inefficiencies (e.g., converting random DRAM accesses to streaming accesses, selectively eliding SRAM bank conflicts), and the mitigate the accuracy loss through approximate-aware network retraining.


\section{Acknowledgements}

The work was supported, in part, by NSF under grants \#2044963 (and its REU supplements) and \#2126642.


\bibliographystyle{IEEEtranS}
\interlinepenalty=10000
\bibliography{refs}

\end{document}